\documentclass[aps,prd,showpacs,onecolumn,preprintnumbers,notitlepage,
nofootinbib,amsfont,amssymb,10pt,singlespacing] {revtex4-1}
\usepackage{amsmath}
\usepackage{times}
\usepackage{braket}
\usepackage{color,graphicx}
\usepackage{slashed}
\usepackage{hyperref}

\newcommand{\beq}{\begin{eqnarray}}
\newcommand{\eeq}{\end{eqnarray}}
\newcommand{\non}{\nonumber\\ }

\newcommand{\psl}{ P \hspace{-2.8truemm}/ }
\newcommand{\nsl}{ n \hspace{-2.2truemm}/ }
\newcommand{\vsl}{ v \hspace{-2.2truemm}/ }

\def\lsim{ {\ \lower-1.2pt\vbox{\hbox{\rlap{$<$}\lower6pt\vbox{\hbox{$\sim$}
}}}\ } }
\def\gsim{ {\ \lower-1.2pt\vbox{\hbox{\rlap{$>$}\lower6pt\vbox{\hbox{$\sim$}
}}}\ } }

\def \jhep{ J. High Energy Phys.  }

\definecolor{Red}{rgb}{1.,0.,0.}

\definecolor{Blue}{rgb}{0.,0.,1.}

\definecolor{nicered}{rgb}{0.7,0.1,0.1}
\definecolor{nicegreen}{rgb}{0.1,0.5,0.1}

\hypersetup{colorlinks,citecolor=nicegreen,linkcolor=nicered}

\begin{document}
%\begin{CJK*}{GB}{gbsn}

\title{\boldmath
Phenomenological studies on the $B_{d,s}^0 \to J/\psi f_0(500) [f_0(980)]$ decays }
%%%==================================================================

\author{Xin~Liu}
%\email[\Gre{Electronic address:} ]{liuxin@jsnu.edu.cn}
%\orcid{0000-0001-9419-7462}
\affiliation{ Department of Physics, Jiangsu Normal University, Xuzhou 221116, China}

\author{Zhi-Tian~Zou}
%\email[\Gre{Electronic address:} ]{zouzt@ytu.edu.cn}
\author{Ying~Li}
%\email[\Gre{Electronic address:} ]{liying@ytu.edu.cn}
\affiliation{ Department of Physics, Yantai University, Yantai 264005,
 China}

\author{Zhen-Jun~Xiao}
%\email[\Gre{Electronic address:} ]{xiaozhenjun@njnu.edu.cn}
\affiliation{ Department of Physics and Institute of Theoretical
Physics, Nanjing Normal University, Nanjing 210023,
 China}

\date{\today}
%%%%%%%%%%%%%%%%%%%%%%%%%%%%%%%%%%%%%%%%%%%%%%%%%%%%%%%%%%%%%%%%%%
\begin{abstract}
Encouraged by the global agreement between theoretical predictions
and experimental measurements for $B \to J/\psi V$ decays, we
extend that perturbative QCD formalism to
$B_{d,s}^0 \to J/\psi f_0(500) [f_0(980)]$ decays at the
presently known next-to-leading order in the quark-antiquark
description of $f_0(500)$ and $f_0(980)$.
With the angle $\phi_f \approx 25^\circ$
of the $f_0(500)-f_0(980)$ mixing in the quark-flavor
basis, we find that the branching ratios of the
$B_d^0 \to J/\psi f_0(500) (\to \pi^+ \pi^-)$ and
$B_{d,s}^0 \to J/\psi f_0(980) (\to \pi^+ \pi^-)$ modes
generally agree with the current data or the
upper limits within uncertainties, except for the seemingly
challenging $B_s^0 \to J/\psi f_0(500) (\to \pi^+\pi^-)$ one.
Then, we further explore the relevant observables of the
$B_{d,s}^0 \to J/\psi f_0(500) [f_0(980)]$
decays, which could provide further constraints on the
mixing angle $\phi_f$ and/or SU(3) flavor symmetry breaking
effects. As a byproduct, we predict
${\rm BR}(B_{d}^0 \to J/\psi f_0(980)(\to K^+ K^-))=5.8^{+3.1}_{-2.9}
\times 10^{-7}$ and ${\rm BR}(B_{s}^0 \to J/\psi f_0(980)(\to K^+ K^-))
=4.6^{+2.6}_{-2.3} \times 10^{-5}$.
All theoretical predictions await the future examinations
with high precision.
\end{abstract}
%%%%%%%%%%%%%%%%%%%%%%%%%%%%%%%%%%%%%%%%%%%%%%%%%%%%%%%%%%%%%%

\preprint{  %{\large J}IANGSU {\large N}ORMAL {\large U}NIVERSITY \hspace{2.4cm}
{\footnotesize JSNU-HEP-2019}}
\maketitle

%
%%%
%%%%%%%%%%%%%%%%% I. INTRODUCTION %%%%%%%%%%%%%%%%%%%%%%%%%%%%%%%%
%%%
%

\section{Introduction}

It is well known that the golden modes $B_d^0 \to J/\psi K_S$ and $B_s^0 \to
J/\psi \phi$ in the heavy $b$ flavor sector provide an ideal ground to test
the standard model(SM) and search for the possible new physics beyond SM.
%Due to
Because of the expected small penguin pollution, the above two decays can
usually offer good opportunities to extract the weak phases $\phi_d$ and $\phi_s$
[or the Cabibbo-Kobayashi-Maskawa(CKM) angles $\beta_d$ and $\beta_s$]
from the indirect {\it CP}-violating asymmetries in the neutral
$B_d^0-\bar B_d^0$ and $B_s^0-\bar B_s^0$ mixings, respectively.
Note that the significant nonzero deviations experimentally
to the SM predictions for the interesting $\sin\phi_d$ and
$\sin\phi_s$ would indicate the exotic new physics beyond SM,
and especially the latter one
is of great interest. However, it is stressed that
the $B_s^0 \to J/\psi \phi$ final state
contains two vector mesons,
which lead to a mixture of {\it CP}-even and {\it CP}-odd
eigenstates; then a complicated angular decomposition
is required to analyze the relevant observables.
Consequently, the extraction of the $B_s^0-\bar B_s^0$ mixing phase
$\phi_s$ suffers from large errors.
Therefore, some new alternative channels are proposed and, in particular,
the $B_s^0 \to J/\psi f_0(980)$ [For simplicity,
$f_0(980)$ is
%will be
abbreviated as $f_0$ in the following context unless otherwise stated.]
is believed to have the  supplementary power to
%in
significantly reduce
%reducing
the error
of $\phi_s$~\cite{Stone:2008ak,Stone:2009hd,Stone:2010dp}.
The underlying reason is that $f_0$ is a $0^{++}$ scalar state
[for example, see the minireview on scalar mesons coming from the Particle Data Group(PDG) in~\cite{Tanabashi:2018oca}],
and thus the final state $J/\psi f_0$ is a {\it CP} eigenstate, which means that, relative to the $B_s^0 \to J/\psi \phi$
channel, there are no needs to perform an angular analysis, and therefore the relevant analysis is simplified  greatly. Indeed, this point has  been proven in
the relevant measurements, for example, the latest one in Ref.~\cite{Aaij:2019mhf}.

Presently, this alternative channel $B_s^0 \to J/\psi f_0$ has been searched through the
resonant contribution with $f_0 \to \pi^+ \pi^-$ by a variety of groups experimentally.
Meanwhile, the expected mixing partner $f_0(500)$, like $\eta-\eta'$ mixing in the pseudoscalar sector, was examined in the
$B_d^0 \to J/\psi f_0(500)$ decay
%(Hereafter,
[hereafter, $f_0(500)$ is
%will be
denoted as $\sigma$ for convenience.]  by the Large Hadron Collider beauty(LHCb)
Collaboration also through resonance studies~\cite{Aaij:2013zpt,Aaij:2014siy}.
The available measurements of branching ratios for the considered $B_d^0 \to J/\psi \sigma$ and $B_s^0 \to J/\psi f_0$ decays are as
follows~\cite{Aaij:2014siy,Li:2011pg,Aaltonen:2011nk,Tanabashi:2018oca},
\beq
{\rm BR}(B_d^0 \to J/\psi f_0(500), f_0 \to \pi^+ \pi^-)  &=&
8.8^{+1.2}_{-1.6}
%\pm 0.5^{+1.1}_{-1.5}
\times 10^{-6}\;,
\label{eq:brdpsis0-ex}\\
{\rm BR}(B_s^0 \to J/\psi f_0(980), f_0 \to \pi^+ \pi^-) &=&
1.28^{+0.18}_{-0.18}\times 10^{-4}\;.
\label{eq:brspsif0-ex}
\eeq
The precision of relevant measurements will be rapidly improved along with
more and more data samples collected at the LHCb and/or Belle-II experiments
in the near future.
Moreover, the upper limits for ${\rm BR}(B_d^0 \to J/\psi f_0)$ and
${\rm BR}(B_s^0 \to J/\psi \sigma)$ are also made currently by
the LHCb Collaboration as follows~\cite{Aaij:2013zpt,Aaij:2014emv}:
\beq
{\rm BR}(B_d^0 \to J/\psi f_0(980), f_0 \to \pi^+ \pi^-) &=& 6.1^{+3.5}_{-2.4}
%^{+3.1+1.7}_{-2.0-1.4}
\times 10^{-7} < 1.1 \times  10^{-6}
\;, \label{eq:brdpsif0-ex}
\\
{\rm BR}(B_s^0 \to J/\psi f_0(500), f_0 \to \pi^+ \pi^-) &<&
4 \times  10^{-6}\;.
\label{eq:brspsis0-ex}
\eeq
It is necessary to stress that the LHCb results  for $B_s^0$ decays
correspond to the time-integrated quantities,
while theory predictions
refer to the branching fractions at $t=0$~\cite{DeBruyn:2012wj},
and may differ by $10\%$.

Furthermore, an interesting ratio $R_{f_0/\phi}$ between the
branching ratios of the alternative $B_s^0 \to J/\psi f_0$
and the golden $B_s^0 \to J/\psi \phi$ channels is defined as~\cite{Stone:2008ak}
\beq
R_{f_0/\phi}&\equiv& \frac{{\rm BR}(B_s^0 \to J/\psi f_0, f_0 \to \pi^+ \pi^-)}
{{\rm BR}(B_s^0 \to J/\psi \phi, \phi \to K^+ K^-)}\;,
\label{eq:rf0phi-def}
\eeq
which has been measured by various groups and the
related results are collected as the following
\cite{Aaij:2011fx,Aaltonen:2011nk,Abazov:2011hv,Khachatryan:2015lua,Amhis:2016xyh},
\beq
R_{f_0/\phi}&\equiv& \frac{{\rm BR}(B_s^0 \to J/\psi f_0, f_0 \to \pi^+ \pi^-)}
{{\rm BR}(B_s^0 \to J/\psi \phi, \phi \to K^+ K^-)}
=\left\{ \begin{array}{lllll}
0.252^{+0.053}_{-0.046}
%^{+0.046+0.027}_{-0.032-0.033}
\;\;\;\;\;({\rm LHCb}),&  \\
0.257^{+0.024}_{-0.024}
%^{+0.020+0.014}_{-0.020-0.014}
\;\;\;\;\;({\rm CDF}),&  \\
0.275^{+0.073}_{-0.073}
%^{+0.041+0.061}_{-0.041-0.061}
\;\;\;\;\;({\rm D0}),&  \\
0.140^{+0.024}_{-0.024}
%^{+0.008+0.023}_{-0.008-0.023}
\;\;\;\;\;({\rm CMS}),&  \\
0.207^{+0.016}_{-0.016}
% \pm 0.016
\;\;\;\;\;({\rm HFLAV})\;. &   \\ \end{array} \right.
\label{eq:rf0phi-data}
\eeq
Meanwhile, another ratio between
${\rm BR}(B_s^0 \to J/\psi f_0, f_0 \to \pi^+ \pi^-)$ and
${\rm BR}(B_s^0 \to J/\psi \phi)$ from different groups is read as follows~\cite{Aaltonen:2011nk,Abazov:2011hv,Khachatryan:2015lua,LHCb:2012ae,Tanabashi:2018oca},
\beq
 \frac{{\rm BR}(B_s^0 \to J/\psi f_0, f_0 \to \pi^+ \pi^-)}
{{\rm BR}(B_s^0 \to J/\psi \phi)}&=& \left\{ \begin{array}{llllll}
0.069^{+0.012}_{-0.012} \;\;\;\;\;({\rm CMS}),&  \\
0.139^{+0.026}_{-0.013} \;\;\;\;\;({\rm LHCb}),&  \\
0.135^{+0.036}_{-0.036} \;\;\;\;\;({\rm D0}),&  \\
0.126^{+0.012}_{-0.012} \;\;\;\;\;({\rm CDF}),&  \\
0.119^{+0.013}_{-0.014} \;\;\;\;\;({\rm PDG\; Fit}),&  \\
0.111^{+0.020}_{-0.018} \;\;\;\;\;({\rm PDG\; Average})\;. &   \\ \end{array} \right.
\label{eq:R-f02pphi-pdg}
 \eeq
These data would be helpful to explore the dynamics involved in the
$B_s^0 \to J/\psi f_0$ decay and to identify the inner structure or
the components of the scalar $f_0$ state.

It is believed that light scalars below 1 GeV could play an important role to
help understand the QCD vacuum because of their same quantum numbers
$J^{PC}=0^{++}$~\cite{Wang:2016wpc}. But, it is unfortunate that
the inner structure of these light scalars such as $\sigma$ and $f_0$
is presently hard to
%be well
understood well due to the complicated nonperturbative QCD dynamics.
Therefore, the interpretation of their components is far from
being straightforward and still in controversy; e.g., see reviews~\cite{Godfrey:1998pd,Close:2002zu,Amsler:2004ps,Klempt:2007cp,
Crede:2008vw,Ochs:2013gi,Tanabashi:2018oca}.
Alternatively, however, the production of $\sigma$ and $f_0$ in the
heavy $D_{(s)}$, $B_{(s)}$, even $B_c$ meson decays could provide
another insight into their inner structure.
In particular, the $B_{d,s}^0 \to J/\psi \sigma (f_0)$ decays could be more
favored because they contain few topologies of Feynman diagrams, as well as
the expectantly small penguin pollution. For example, Stone and Zhang ever
suggested in Ref.~\cite{Stone:2013eaa} that these channels could be used
to discern the $q \bar{q}$ or tetraquark nature of scalars, and an upper limit
of the mixing angle between $\sigma$ and $f_0$ was provided with the help of $B_d^0
\to J/\psi \sigma$ and $B_s^0 \to J/\psi f_0$ decays as $29^\circ$ at 90\%
confidence level for the $\sigma$ and $f_0$ being $q\bar q$ states.

On the theoretical side, some of these $B_{d,s}^0 \to J/\psi \sigma (f_0)$ modes
have been investigated to a different extent with different methods/approaches in the literature
\cite{Colangelo:2010bg,Colangelo:2010wg,Leitner:2010fq,Fleischer:2011au,
Li:2012sw,Liang:2014tia,Bayar:2014qha,Wang:2015uea,Close:2015rza,
Wang:2016wpc,Daub:2015xja,Ropertz:2018stk}, and,
in particular,
\begin{itemize}
\item[(a)]
Colangelo {\it et al.} studied the $B_s^0 \to J/\psi f_0$ decay by using the
light-cone QCD sum rule and factorization assumption in
Ref.~\cite{Colangelo:2010bg} with leading order prediction ${\rm BR}
(B_s^0 \to J/\psi f_0)= 3.1 \pm 2.4 \times 10^{-4}$
and the next-to-leading order(NLO) one ${\rm BR}(B_s^0 \to J/\psi f_0)=
5.3 \pm 3.9 \times 10^{-4}$, and using generalized factorization
and SU(3) flavor symmetry in Ref.~\cite{Colangelo:2010wg} with different
branching ratios $4.7 \pm 1.9 \times 10^{-4}$ and $2.0 \pm 0.8
\times 10^{-4}$, respectively. Notice that here
$f_0$ was assumed as a pure $s\bar s$ state.

\item[(b)]
By assuming $f_0$ as an $s\bar s$ state, Leitner {\it et al.} estimated
the $B_s^0 \to J/\psi f_0$ decay rate around $5.0 \times 10^{-4}$ in the
QCD factorization approach~\cite{Leitner:2010fq}, based on
reproduction of the data about ${\rm BR}(B_s^0 \to J/\psi \phi)$.

\item[(c)]
Fleischer {\it et al.} showed the anatomy of $B_{d,s}^0 \to J/\psi f_0$
in Ref.~\cite{Fleischer:2011au} by considering the $q\bar q$ and
tetraquark pictures of the $f_0$ state. And they obtained the branching
ratios with different mixing angles $\varphi_M$ in the conventional
two-quark picture:
${\rm BR}(B_s^0 \to J/\psi f_0)|_{\varphi_M=0^\circ} \simeq 1.9 \times 10^{-4}$
and ${\rm BR}(B_s^0 \to J/\psi f_0)|_{\varphi_M=41.6^\circ} \simeq 4.8 \times
10^{-4}$ by using factorization approximation and SU(3) flavor symmetry.
Meanwhile, the $B_d^0 \to J/\psi f_0(\to \pi^+ \pi^-)$ decay rate
$\sim 1.65^{+0.34}_{-0.29} \times 10^{-6}$ was also predicted.

\item[(d)]
Under the assumption of two-quark structure and the $\sigma-f_0$ mixing,
Li {\it et al.} studied the $B_s^0 \to J/\psi \sigma (f_0)$ decays with
a mixed ``QCD factorization plus perturbative QCD(PQCD) factorization" approach~\cite{Li:2012sw} and predicted the
branching ratios ${\rm BR}(B_s^0 \to J/\psi f_0)=2.43^{+0.30}_{-0.31} \times
10^{-4}$ and ${\rm BR}(B_s^0 \to J/\psi \sigma)= 4.72^{+0.62}_{-0.59} \times
10^{-5}$, corresponding to the mixing angle $\phi_f$ about $\pm 34^\circ$.
\end{itemize}

In light of the current measurements on various observables performed
by the LHCb Collaboration with good
precision, it is essential to make
a systematic investigation on all of the $B_{d,s}^0 \to J/\psi \sigma (f_0)$
modes. Encouraged by the global agreement between the data and the
theoretical predictions in the PQCD approach
~\cite{Keum:2000ph,Keum:2000wi,Lu:2000em,Lu:2000hj} on the $B \to J/\psi V$
decays at the NLO accuracy~\cite{Liu:2013nea}, we
%will
extend that
formalism to the $B_{d,s}^0 \to J/\psi \sigma (f_0)$ decays
in the quark-antiquark description of $\sigma$ and $f_0$
with including the known NLO corrections in $\alpha_s$, namely, the vertex
corrections. It is well known that, as one of the popular factorization
methods based on QCD dynamics, the PQCD approach has been widely employed
to calculate the hadronic matrix elements in the nonleptonic decays
of heavy $b$ quark mesons.
%Due to
Because of the introduction of the Sudakov factors arising from $k_T$ resummation~\cite{Botts:1989kf,Li:1992nu} and threshold
resummation~\cite{Li:2001ay,Li:2002mi}, respectively,
the PQCD approach could be utilized to compute the nonfactorizable
emission and the annihilation diagrams safely, apart from the factorizable emission ones.
With the perturbative calculations of both tree and penguin
amplitudes in the PQCD approach, we could provide the predictions on the
observables such as  the {\it CP}-averaged branching ratios, the {\it CP}-violating
asymmetries, and so forth with much more reliability.
Hence, these reliable calculations would help us to further investigate the
impact of the penguin contributions to the {\it CP} asymmetry measurements,
even the extraction of weak phases $\phi_{d,s}$, and
explore the useful information such as the mixing angle $\phi_f$ between the mixtures of $\sigma$ and $f_0$, if they are really the $q\bar q$ mesons.

The rest of this paper is organized as follows: After this introduction,
Sec.~\ref{sec:form} is devoted to the analysis of
decay amplitudes for the $B_{d,s}^0 \to J/\psi \sigma(f_0)$ modes
in the PQCD approach. The essential nonperturbative inputs
are also collected in this section. The numerical results and
phenomenological analyses for the {\it CP}-averaged
branching ratios,
{\it CP}-violating asymmetries, and other interesting observables
of the considered decays
are given in Sec.~\ref{sec:r&d}. As a byproduct, we also present
the {\it CP}-averaged branching ratios of
$B_{d,s}^0 \to J/\psi f_0(\to K^+ K^-)$ decays
in this section.
We summarize this work and conclude in Sec.~\ref{sec:summary}.

%%%
%%%%%%%%%%%%%%%%% II. Formalism %%%%%%%%%%%%%%%%%%%%%%%%%%%%%%%%
%%%
%
\section{\boldmath Decay amplitudes of $B_{d,s}^0 \to J/\psi \sigma(f_0)$ and Essential inputs}\label{sec:form}

%%%%=============================================================
\begin{figure}[!!htb]
  \centering
  \begin{tabular}{l}
  \includegraphics[width=0.8\textwidth]{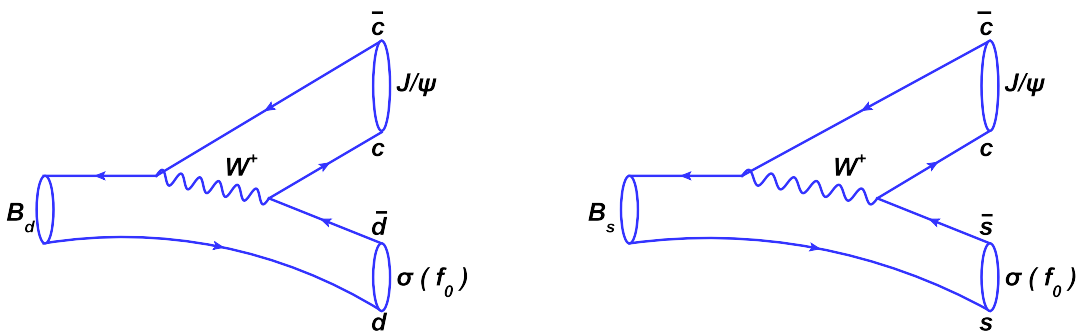}
  \end{tabular}
  \caption{ Leading quark-level Feynman diagrams contributing to the $B_{d,s}^0 \to J/\psi \sigma (f_0)$ decays. }
  \label{fig:fig1}
\end{figure}
%%%%==============================================================

Similar to $B_{d,s}^0 \to J/\psi \eta (\eta^{\prime})$ decays in the
pseudoscalar sector~\cite{Liu:2012ib}, the leading quark-level Feynman
diagrams contributing to the $B_{d,s}^0 \to J/\psi \sigma (f_0)$ decays
have been illustrated in Fig.~\ref{fig:fig1}.
Before writing down the decay amplitudes of the considered $B_{d,s}^0
\to J/\psi \sigma (f_0)$ channels, it is essential to make some remarks
on the mixing between $\sigma$ and $f_0$.
Analogous to the $\eta-\eta'$ mixing,
this scalar $\sigma-f_0$ mixing can also be described by
a $2\times 2$ rotation matrix with a single angle $\phi_{f}$
in the quark-flavor basis, namely,
 \beq
\left(
\begin{array}{c} \sigma \\ f_0 \\ \end{array} \right ) &=&
  \left( \begin{array}{cc}
 \cos{\phi_{f}} & -\sin{\phi_{f}} \\
 \sin{\phi_{f}} & \cos{\phi_{f}} \end{array} \right )
 \left( \begin{array}{c}  f_q\\ f_s \\ \end{array} \right )\;.
 \label{eq:mix-s0-f0}
 \eeq
with the quark-flavor states
$f_q \equiv \frac{u\bar u + d\bar d}{\sqrt{2}}$
and $f_s\equiv s\bar s$.
Various mixing angle $\phi_f$ measurements have been derived and summarized in the
literature with a wide range of values; for example,
see Refs.~\cite{Cheng:2002ai,Cheng:2005nb,Fleischer:2011au,Cheng:2013fba}.
However, it is worth of pointing out that, based on the recent
measurement and the accompanied discussion performed
by the LHCb Collaboration~\cite{Aaij:2013zpt}, the upper limits $|\phi_f|<31^\circ$
have been set for the first time in the $B$ meson decays with a two-quark structure description of $\sigma$ and $f_0$.
Therefore, in other words, the agreement of
{\it CP}-averaged branching ratios
for the $B_{d,s}^0 \to J/\psi \sigma(f_0)$ decays between the
experimental measurements and the PQCD predictions in this work is expected
to provide some useful information to further constrain the possible
range of this $\phi_f$ angle.

According to the aforementioned mixing pattern, the $B_{d,s}^0 \to J/\psi \sigma(f_0)$
decay amplitudes could then be written explicitly with the help of $B_{d(s)}^0 \to
J/\psi f_{q(s)}$ as follows,
\beq
{\cal A}(B_d^0 \to J/\psi \sigma) &=&
{\cal A}(B_d^0 \to J/\psi f_q)\cdot \cos\phi_f\;,
\label{eq:decay-dps0}\\
{\cal A}(B_d^0 \to J/\psi f_0) &=&
{\cal A}(B_d^0 \to J/\psi f_q)\cdot \sin\phi_f\;;
\label{eq:decay-dpf0}\\
{\cal A}(B_s^0 \to J/\psi \sigma) &=&
{\cal A}(B_s^0 \to J/\psi f_s)\cdot (-\sin\phi_f)\;,
\label{eq:decay-sps0}\\
{\cal A}(B_s^0 \to J/\psi f_0) &=&
{\cal A}(B_s^0 \to J/\psi f_s)\cdot \cos\phi_f\;,
\label{eq:decay-spf0}
\eeq
which yield the following relations:
\beq
|{\cal A}(B_d^0\to J/\psi \sigma)|^2 +|{\cal A}(B_d^0\to J/\psi f_0)|^2
&=& |{\cal A}(B_d^0 \to J/\psi f_q)|^2\;, \\
|{\cal A}(B_s^0\to J/\psi \sigma)|^2 +|{\cal A}(B_s^0\to J/\psi f_0)|^2
&=& |{\cal A}(B_s^0 \to J/\psi f_s)|^2\;.
\eeq
Here, the decay amplitudes of $B_{d(s)}^0$ decaying into the flavor state
$f_{q(s)}$ could be easily obtained from those in
the $B_{d(s)}^0 \to J/\psi \omega(\phi)$ modes correspondingly
in the PQCD approach, which is
%will be
clarified later.
These formulas indicate that the theoretically reliable estimates of the
perturbative and nonperturbative
QCD dynamics in the $B_{d(s)}^0 \to J/\psi f_{q(s)}$ modes
are very important to understand the $B_{d,s}^0 \to J/\psi \sigma (f_0)$
decays experimentally, and vice versa. It is worth
%y of
mentioning that
the wave functions associated with
light-cone distribution amplitudes
that describe the hadronization of valence quark and valence antiquark
in a meson are the only nonperturbative inputs in the PQCD calculations
and are processes independent. It is fortunate that the
nonperturbative
QCD dynamics of the above-mentioned initial and final hadrons has
been investigated in the literature.
\begin{itemize}
\item[(a)]
It is remarked that the $B\to J/\psi P(V)$ decays[$P(V)$ stands for the light
pseudoscalar(vector) mesons] have been studied in
the PQCD approach at the NLO
accuracy~\cite{Chen:2005ht,Li:2006vq,Liu:2010zh,Liu:2012ib,Liu:2013nea,Liu:2014doa}
with the same wave functions and distribution amplitudes for the heavy
$B_{d,s}^0$ and $J/\psi$ mesons. Furthermore, the general consistency
between theory and experiment in the SM for the branching ratios
of those considered decays has been obtained. Thus, in this work,
we
%will
adopt the same wave functions and distribution amplitudes of
$B_{d,s}^0$ and $J/\psi$ as those used in,
for example, Ref.~\cite{Liu:2013nea} and references therein, as well as
the relevant hadronic parameters.

\item[(b)]
For the scalar flavor states $f_q$ and $f_s$, the light-cone wave function
can generally be defined as~\cite{Cheng:2005ye}
\beq
 \Phi_{f_{q(s)}}(x) &=&  \frac{i}{\sqrt{2 N_c}}
 \Biggl\{\psl\, \phi_{f_{q(s)}}(x) +
 m_{f_{q(s)}}\, \phi_{f_{q(s)}}^S(x) + m_{f_{q(s)}}
 (\nsl \vsl - 1)\, \phi_{f_{q(s)}}^T(x) \Biggr\}_{\alpha\beta}\;,
 \label{eq:wf-fqs}
\eeq
where $N_c$, $\phi_{f_{q(s)}}$, and $\phi_{f_{q(s)}}^{S,T}$, $m_{f_{q(s)}}$,
$n$, and $v$, and $\alpha,\beta$ are the color factor, the leading twist,
and twist 3 distribution amplitudes, the mass of $f_{q(s)}$,
the dimensionless lightlike unit vectors $n=(1,0,{\bf 0}_T)$ and
$v=(0,1,{\bf 0}_T)$, and the color indices, respectively, while $x$
denotes the momentum fraction carried by the quark in the meson.

The light-cone distribution amplitudes up to twist 3 as shown
in Eq.~(\ref{eq:wf-fqs}) have been investigated in the QCD sum rule
technique\footnote{%Due to
Because of
charge conjugation invariance
or conservation of vector current, the neutral scalar $\sigma$ and $f_0$
mesons cannot be produced through the vector current, which, consequently,
results in the zero values of their vector decay constants,
i.e., $f_{f_q} = f_{f_s} = 0$.}\cite{Cheng:2005ye}
with the contributions arising from only the odd Gegenbauer polynomials,
\beq
\phi_{f_{q(s)}}&=& \frac{\bar f_{f_{q(s)}}(\mu)}{2\sqrt{2 N_c}}
\Bigg\{6x(1-x)\Bigg[B_1^{q(s)}(\mu) C_1^{3/2}(2x-1)
+B_3^{q(s)}(\mu) C_3^{3/2}(2x-1)\Bigg]\Bigg\}\;,
\eeq
\beq
\phi_{f_{q(s)}}^S &=&\frac{1}{2\sqrt {2N_c}}\bar f_{f_{q(s)}}(\mu)\;,
\,\,\,\,\,\,\,\qquad
\phi_{f_{q(s)}}^T= \frac{1}{2\sqrt {2N_c}}\bar f_{f_{q(s)}}(\mu)(1-2x)\;,
\eeq
where the scalar decay constants $\bar f_{f_{q}}(\mu)$
and $\bar f_{f_s}(\mu)$ and the Gegebnbauer moments
$B_{1,3}^{q(s)}(\mu)$ at the normalization scale $\mu =1$ GeV
are as follows~\cite{Cheng:2005ye}:
\beq
 \bar f_{f_q}  &\simeq& 0.35\;{\rm  GeV}\;, \qquad
  \bar f_{f_s}  \simeq 0.33\;{\rm  GeV}\;;
  \label{eq:dc-s}
\eeq
\beq
B_1^q &=& -0.92 \pm 0.08\;, \qquad B_3^q = -1.00 \pm 0.05\;,
\qquad
B_{1,3}^s \simeq 0.8 B_{1,3}^q\;.
\label{eq:gm-s}
\eeq
The expressions for the Gegenbauer polynomials $C_1^{3/2}(t)$ and $C_3^{3/2}(t)$
can be found explicitly, for example, from Eqs.~(A8) and~(A10) in Ref.~\cite{Li:2006jv} with $\lambda=3/2$.
\end{itemize}

The related weak effective Hamiltonian $H_{{\rm eff}}$ for
the $B_{d(s)}^0 \to J/\psi f_{q(s)}$ decays mentioned above
can be written as~\cite{Buchalla:1995vs}
\beq
H_{\rm eff}\, &=&\, {G_F\over\sqrt{2}}
\biggl\{ V^*_{cb}V_{cQ} [ C_1(\mu)O_1^{c}(\mu)
+C_2(\mu)O_2^{c}(\mu) ]
 - V^*_{tb}V_{tQ} [ \sum_{i=3}^{10}C_i(\mu)O_i(\mu) ] \biggr\}+ {\rm h.c.}\;,
\label{eq:heff}
\eeq
with the Fermi constant $G_F=1.16639\times 10^{-5}{\rm
GeV}^{-2}$, the light $Q = d, s$ quark,
and Wilson coefficients $C_i(\mu)$ at the renormalization scale
$\mu$. The local four-quark
operators $O_i(i=1,\cdots,10)$ are written as
\begin{enumerate}
\item[]{(1) current-current(tree) operators}
\begin{eqnarray}
{\renewcommand\arraystretch{1.5}
\begin{array}{ll}
\displaystyle
O_1^{c}\, =\,
(\bar{Q}_\alpha c_\beta)_{V-A}(\bar{c}_\beta b_\alpha)_{V-A}\;,
& \displaystyle
O_2^{c}\, =\, (\bar{Q}_\alpha c_\alpha)_{V-A}(\bar{c}_\beta b_\beta)_{V-A}\;;
\end{array}}
\label{eq:operators-1}
\end{eqnarray}

\item[]{(2) QCD penguin operators}
\begin{eqnarray}
{\renewcommand\arraystretch{1.5}
\begin{array}{ll}
\displaystyle
O_3\, =\, (\bar{Q}_\alpha b_\alpha)_{V-A}
\sum_{q'}(\bar{q}'_\beta q'_\beta)_{V-A}\;,
& \displaystyle
O_4\, =\, (\bar{Q}_\alpha b_\beta)_{V-A}
\sum_{q'}(\bar{q}'_\beta q'_\alpha)_{V-A}\;,
\\
\displaystyle
O_5\, =\, (\bar{Q}_\alpha b_\alpha)_{V-A}
\sum_{q'}(\bar{q}'_\beta q'_\beta)_{V+A}\;,
& \displaystyle
O_6\, =\, (\bar{Q}_\alpha b_\beta)_{V-A}
\sum_{q'}(\bar{q}'_\beta q'_\alpha)_{V+A}\;;
\end{array}}
\label{eq:operators-2}
\end{eqnarray}

\item[]{(3) electroweak penguin operators}
\begin{eqnarray}
{\renewcommand\arraystretch{1.5}
\begin{array}{ll}
\displaystyle
O_7\, =\,
\frac{3}{2}(\bar{Q}_\alpha b_\alpha)_{V-A}
\sum_{q'}e_{q'}(\bar{q}'_\beta q'_\beta)_{V+A}\;,
& \displaystyle
O_8\, =\,
\frac{3}{2}(\bar{Q}_\alpha b_\beta)_{V-A}
\sum_{q'}e_{q'}(\bar{q}'_\beta q'_\alpha)_{V+A}\;,
\\
\displaystyle
O_9\, =\,
\frac{3}{2}(\bar{Q}_\alpha b_\alpha)_{V-A}
\sum_{q'}e_{q'}(\bar{q}'_\beta q'_\beta)_{V-A}\;,
& \displaystyle
O_{10}\, =\,
\frac{3}{2}(\bar{Q}_\alpha b_\beta)_{V-A}
\sum_{q'}e_{q'}(\bar{q}'_\beta q'_\alpha)_{V-A}\;,
\end{array}}
\label{eq:operators-3}
\end{eqnarray}
\end{enumerate}
with the notations $(\bar{q}'q')_{V\pm A} = \bar q' \gamma_\mu (1\pm \gamma_5)q'$.
The index $q'$ in the summation of the above operators runs
through $u,\;d,\;s$, $c$, and $b$. The standard combinations $a_i$ of
Wilson coefficients are defined as follows,
  \beq
a_1&=& C_2 + \frac{C_1}{3}\;, \qquad  a_2 = C_1 + \frac{C_2}{3}\;,\non
 a_i &=& C_i + \frac{C_{i \pm 1}}{3}(i=3 - 10) \;,
  \eeq
where the upper(lower) sign applies, when $i$ is odd(even). It should
be mentioned that, similar to $B \to J/\psi V$ decays~\cite{Liu:2013nea},
the NLO Wilson coefficients $C_i(i=1,\cdots,10)$ and the strong coupling
constant $\alpha_s$ at two-loop level with $\Lambda_{\rm QCD}^{(5)}=0.225$
GeV~\cite{Buchalla:1995vs} are
%will be
adopted in the calculations of the
$B_{d(s)}^0 \to J/\psi f_{q(s)}$ decay amplitudes.

As for the decay amplitudes of $B_{d(s)}^0 \to J/\psi f_{q(s)}$, we adopt
$F_{fe}$ and $M_{nfe}$ to stand for the contributions of factorizable
emission and nonfactorizable emission diagrams from $(V-A)(V-A)$
operators. The explicit expressions of these two Feynman amplitudes
$F_{fe}$ and $M_{nfe}$ can be obtained
by replacing the distribution amplitudes $\phi_{\omega(\phi)}$
and $\phi_{\omega(\phi)}^{s,t}$ in the
$B_{d(s)}^0 \to [J/\psi \omega(\phi)]_L$ mode($L$ stands
for longitudinal polarization),
i.e., Eqs.~(37) and~(40) in~\cite{Liu:2013nea}, with those
$\phi_{f_{q(s)}}$ and $\phi_{f_{q(s)}}^{S,T}$ correspondingly.
Meanwhile, the masses of the light mesons should be replaced
correspondingly too. Therefore, for simplicity, we do
%will
not present the factorization formulas of $F_{fe}$ and $M_{nfe}$
for the $B_{d(s)} \to J/\psi f_{q(s)}$ decays
in this work. The readers can refer to Ref.~\cite{Liu:2013nea}
for detail.

By taking various contributions from the relevant Feynman diagrams
into consideration, the total decay amplitudes for  $B_{d(s)}^0 \to
J/\psi f_{q(s)}$ channels are given as
\beq
{\cal A}(B_{d(s)}^0 \to J/\psi f_{q(s)}) &=& F_{fe}
f_{J/\psi} \Bigg\{ V_{cb}^*V_{cd(s)}\; \tilde{a}_2 -V_{tb}^{*}V_{td(s)} \Bigg (
\tilde{a}_3+\tilde{a}_5+\tilde{a}_7 +\tilde{a}_9 \Bigg) \Bigg\} \non &&
 + M_{nfe}  \Bigg\{V_{cb}^*V_{cd(s)} C_2
 - V_{tb}^*V_{td(s)}  \Bigg (C_4-C_6-C_8+C_{10}\Bigg) \Bigg\}\;,
 \label{eq:tda-bds2psif}
\eeq
where $\tilde{a}_i$ stands for the effective Wilson coefficients that
include the contributions arising from the
vertex corrections at NLO
level. The explicit expressions of $\tilde{a}_i$ can be found
in Appendix~\ref{sec:app1}.

%%%%%%%%%%%%%%%%%%%%%%%%%%%%%%%%%%%%%%%%%%%%%%%%%%%%%%%%%%%%%%%%
\section{Numerical Results and Discussions} \label{sec:r&d}
%%%%%%%%%%%%%%%%%%%%%%%%%%%%%%%%%%%%%%%%%%%%%%%%%%%%%%%%%%%%%%%%

%%%%%%%%%%%%%%%%%%%%%%%%%%%%%%%%
%\subsection{Input quantities}
%%%%%%%%%%%%%%%%%%%%%%%%%%%%%%%%
We present the theoretical predictions about the interesting
observables such as {\it CP}-averaged branching ratios and
{\it CP}-violating asymmetries for those
considered $B_{d,s}^0 \to J/\psi \sigma(f_0)$ decay modes in the PQCD approach.
In numerical calculations, central values of the input parameters are
%will be
used implicitly unless otherwise stated.

The masses~(in units of {\rm GeV}) and $B_{d,s}^0$ meson lifetime(in {\rm ps})
are taken from Refs.~\cite{Cheng:2005ye,Tanabashi:2018oca},
\beq
m_W &=& 80.41\;, \quad  m_{B}= 5.28\;, \quad
m_{B_s} = 5.37 \;, \quad  m_b = 4.8\;, \quad
m_{f_q} = 0.99\;; \non   m_{f_s} &=& 1.02\;,
\quad m_c = 1.5\;, \quad m_{J/\psi}= 3.097\;,
\quad \tau_{B_d}= 1.520\;, \quad  \tau_{B_s}= 1.509\;.
\label{eq:mass}
\eeq

For the CKM matrix elements, we adopt the Wolfenstein
parametrization up to corrections of ${\cal O}(\lambda^5)$
and the updated parameters $A=0.836$,  $\lambda=0.22453$,
$\bar{\rho}=0.122^{+0.018}_{-0.017}$, and
$\bar{\eta}=0.355^{+0.012}_{-0.011}$~\cite{Tanabashi:2018oca}.

%%%%%%%%%%%%%%%%%%%%%%%%%%%%%%%%%%%%%%%%%%%%%%%%%%%%%%%%%%%%%%%%%%%%%%%%%%%%%%%%%%%%%%%%%%%%%%%%%%%%%%%%%
%\subsection{\boldmath {\it CP}-averaged Branching Ratios}\label{ssec:cp-brs}
%%%%%%%%%%%%%%%%%%%%%%%%%%%%%%%%%%%%%%%%%%%%%%%%%%%%%%%%%%%%%%%%%%%%%%%%%%%%%%%%%%%%%%%%%%%%%%%%%%%%%%%%%

By employing those decay amplitudes, i.e.,
Eqs.~(\ref{eq:decay-dps0})-(\ref{eq:decay-spf0})
and Eq.~(\ref{eq:tda-bds2psif}), the formulas of
branching ratios for
the considered $B_{d,s}^0 \to J/\psi \sigma(f_0)$ decays
can be written as
\beq
{\rm BR}(B_{d,s}^0 \to J/\psi \sigma(f_0))&\equiv&
\tau_{B_{d(s)}^0}\cdot \Gamma(B_{d,s}^0 \to J/\psi \sigma(f_0))
\non &=& \tau_{B_{d(s)}^0}\cdot\frac{G_{F}^{2} m^{7}_{B_{d(s)}^0}}
{16 \pi}\cdot \Phi_{\sigma, f_0}^{d, s} \cdot |{\cal A}(B_{d,s}^0\to
J/\psi \sigma(f_0))/m^{2}_{B_{d(s)}^0}|^2\;,
\label{eq:br-def}
\eeq
where $\tau_{B_{d(s)}^0}$ is the lifetime of $B_{d(s)}^0$ meson
and $\Phi_{\sigma, f_0}^{d,s}$ stands for the phase space factors of
$B_{d,s}^0 \to J/\psi \sigma(f_0)$ decays,
\beq
\Phi_{\sigma(f_0)}^{d} &\equiv&
\Phi(m_{J/\psi}/m_{B_{d}^0}, m_{\sigma(f_0)}/m_{B_{d}^0})\;,
\qquad
\Phi_{\sigma(f_0)}^{s} \equiv
\Phi(m_{J/\psi}/m_{B_{s}^0}, m_{\sigma(f_0)}/m_{B_{s}^0})\;,
\eeq
with $\Phi(x,y) \equiv \sqrt{[1-(x+y)^2] [1-(x-y)^2]}$~\cite{Fleischer:2011au},
$m_\sigma =0.5$ GeV, and $m_{f_0}=0.98$ GeV.

As discussed in the literature, up to now, the mixing angle $\phi_f$
between the mixtures of $\sigma$ and $f_0$ could not be determined definitely yet
and is still in controversy. Various values and/or ranges have been analyzed; e.g.,
see Ref.~\cite{Fleischer:2011au,Cheng:2013fba} and references contained therein.
However, based on lots of measurements via resonance investigations on the
$B_{d,s}^0 \to J/\psi \sigma (f_0)$ decays as presented in
Eqs.~(\ref{eq:brdpsis0-ex})-(\ref{eq:brspsis0-ex}),
it may be more interesting to consider the
dependence of the {\it CP}-averaged branching ratios of
$B_{d,s}^0 \to J/\psi \sigma/f_0(\to \pi^+ \pi^-)$
with the angle $\phi_f$
in the PQCD approach, which would hint effectively at
the acceptable value of $\phi_f$ in this work.
Certainly, different from the corresponding quasi-two-body
decays~\cite{Wang:2015uea}, the $\sigma/f_0 \to \pi^+ \pi^-$ decay
rate is
%will be
regarded as an input in this work.

It is noted that the $f_0$ is an elusive object that decays
largely into $\pi^+ \pi^-$ but also decays into $K^+ K^-$.
By combining the {\it BABAR} measurements about the $B \to KKK, K\pi\pi$ decays
and the BES measurements about $\psi(2S) \to \gamma \chi_{c0}(\to f_0 f_0)$
decays with either both $f_0$ decaying into $\pi^+ \pi^-$ or one into
$\pi^+ \pi^-$ and the other into $K^+ K^-$
pairs~\cite{Aubert:2006nu,Ablikim:2004cg,Ablikim:2005kp,Ecklund:2009aa},
the average of these two measurements could give~\cite{Aaij:2013zpt}
\beq
{\cal R}&\equiv&\frac{{\cal B}(f_0 \to K^+ K^-)}{{\cal B}(f_0 \to \pi^+ \pi^-)}
= 0.35^{+0.15}_{-0.14}\;,
\label{eq:r-f02kkpp-ex}
\eeq
which
%will
results in the following branching ratios explicitly:
\beq
{\cal B}(f_0 \to \pi^+ \pi^-)&=& 0.45^{+0.07}_{-0.05}  \;, \qquad
{\cal B}(f_0 \to K^+ K^-) = 0.16^{+0.04}_{-0.05}\;;
\label{eq:br-pp-kk-f0}
\eeq
by employing the formulas ${\cal B}(f_0 \to \pi^+ \pi^-) = \frac{2}{4{\cal R}+3}$
and ${\cal B}(f_0 \to K^+ K^-) = \frac{2{\cal R}}{4{\cal R}+3}$~\cite{Fleischer:2011au}. Here, the dominance of
$f_0$ decaying into $\pi\pi$ and $KK$ is assumed, and the only other decays
are also assumed to $\pi^0 \pi^0$, half
of the $\pi^+ \pi^-$ rate, and to $K^0 \bar K^0$, taken equal to $K^+ K^-$.
For the $\sigma$ meson, it is assumed that the only decays are into two pions. Then,
following from the isospin Clebsch-Gordan coefficients, the $\sigma \to \pi^+ \pi^-$
decay rate could be obtained as $\frac{2}{3}$. In order to estimate the uncertainties
from $\sigma \to \pi^+ \pi^-$ decay, the variations with $10\%$ of the central value,
i.e., ${\cal B}(\sigma \to \pi^+ \pi^-) \simeq 0.67 \pm 0.07$, are
%will be
taken into account in the following estimations.

%%%%%%%%%%%%%%%%%%%%%%%%%%%%%%%%%%%%%%%%%%%%%%%%%%%%%%%%%%%%%%%%%%%%%%%%%%%%%%%%%%%%%%%%%%%%%%%%%%
\begin{figure}[!htp]
\begin{center}
\hspace{-1 cm}
\includegraphics[scale=0.55]{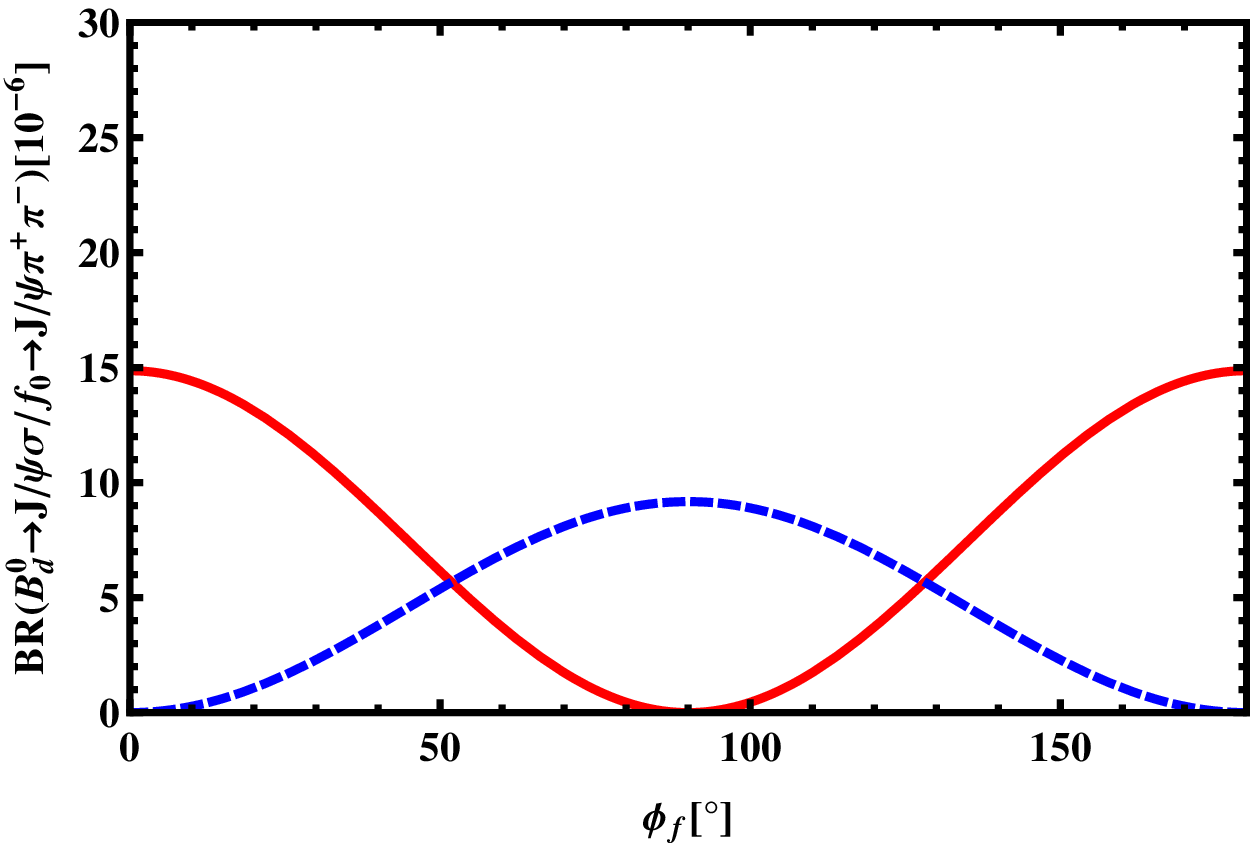}\hspace{0.6cm}
\includegraphics[scale=0.55]{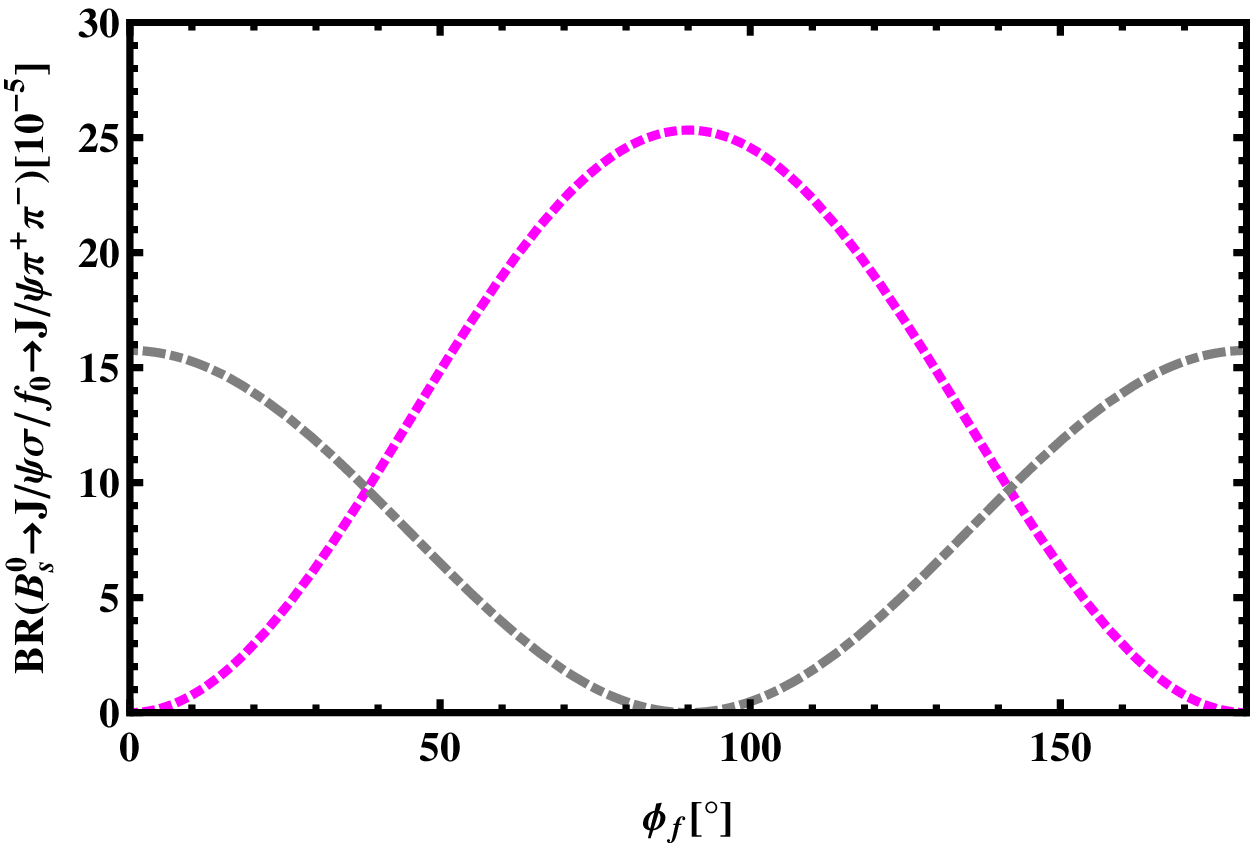}
\vspace{-0.5cm}
\caption{  Dependence on the mixing angle $\phi_{f}$ of the central values
for ${\rm BR}(B_{d,s}^0 \to J/\psi \sigma/f_0 \to J/\psi \pi^+ \pi^-)$ in the PQCD approach:
The red solid\ [blue dashed] line corresponds to the $B_d^0 \to J/\psi \sigma(\to \pi^+ \pi^-)\
[B_d^0 \to J/\psi f_0(\to \pi^+ \pi^-)]$ decay, and the magenta dotted\ [gray dot-dashed] line
corresponds to the $B_s^0 \to J/\psi \sigma(\to \pi^+ \pi^-)\ [B_s^0 \to J/\psi f_0(\to \pi^+ \pi^-)]$
decay, respectively.}
\label{fig:fig2}
\end{center}
\end{figure}
%%%%%%%%%%%%%%%%%%%%%%%%%%%%%%%%%%%%%%%%%%%%%%%%%%%%%%%%%%%%%%%%%%%%%%%%%%%%%%%%%%%%%%%%%%%%%%%%%%

Therefore, armed with
${\cal B}(f_0 \to \pi^+ \pi^-)$ and ${\cal B}(\sigma \to \pi^+ \pi^-)$,
the $B_{d,s}^0 \to J/\psi \sigma/f_0(\to \pi^+ \pi^-)$ decay rates
varying with the mixing angle $\phi_f$ could be further written
theoretically as~\cite{Cheng:2003xc}
\beq
{\rm BR}(B_d^0 \to J/\psi \sigma, \sigma\to \pi^+ \pi^-) &\equiv&
{\rm BR}(B_d^0 \to J/\psi \sigma) {\cal B}(\sigma \to \pi^+ \pi^-)\non
&\propto& \tau_{B_d^0}\cdot \Phi_{\sigma}^d\cdot m_{B_d^0}^7 \cdot|{\cal A}(B_d^0 \to J/\psi f_q)/m_{B_d^0}^2|^2 \cdot \cos^2\phi_f\;,
\label{eq:brdpsis0-pipi-phi}\\
{\rm BR}(B_d^0 \to J/\psi f_0, f_0\to \pi^+ \pi^-) &\equiv&
{\rm BR}(B_d^0 \to J/\psi f_0) {\cal B}(f_0 \to \pi^+ \pi^-)\non
&\propto& \tau_{B_d^0}\cdot \Phi_{f_0}^d\cdot m_{B_d^0}^7 \cdot|{\cal A}(B_d^0 \to J/\psi f_q)/m_{B_d^0}^2|^2 \cdot \sin^2\phi_f\;;
\label{eq:brdpsif0-pipi-phi}
\eeq
\beq
{\rm BR}(B_s^0 \to J/\psi \sigma, \sigma\to \pi^+ \pi^-) &\equiv&
{\rm BR}(B_s^0 \to J/\psi \sigma) {\cal B}(\sigma \to \pi^+ \pi^-)\non
&\propto& \tau_{B_s^0}\cdot \Phi_{\sigma}^s\cdot m_{B_s^0}^7 \cdot|{\cal A}(B_s^0 \to J/\psi f_s)/m_{B_s^0}^2|^2 \cdot \sin^2\phi_f\;,
\label{eq:brspsis0-pipi-phi}\\
{\rm BR}(B_s^0 \to J/\psi f_0, f_0\to \pi^+ \pi^-) &\equiv&
{\rm BR}(B_s^0 \to J/\psi f_0) {\cal B}(f_0 \to \pi^+ \pi^-)\non
&\propto& \tau_{B_s^0}\cdot \Phi_{f_0}^s\cdot m_{B_s^0}^7 \cdot|{\cal A}(B_s^0 \to J/\psi f_s)/m_{B_s^0}^2|^2 \cdot \cos^2\phi_f\;.
\label{eq:brspsif0-pipi-phi}
\eeq

By employing the decay amplitudes and the hadronic inputs, we plot the
{\it CP}-averaged branching ratios in the PQCD approach at the known NLO level
of $B_{d,s}^0 \to J/\psi \sigma/f_0(\to \pi^+ \pi^-)$ decays depending on the
angle $\phi_f$, which can be seen explicitly in
Fig.~\ref{fig:fig2}. Here, the central values of the relevant branching ratios
varying with $\phi_f$ are presented for clarification. By comparing
with the data as shown in Eqs.~(\ref{eq:brdpsis0-ex})-(\ref{eq:brspsis0-ex}),
one can easily observe the overall consistency between experiment and theory
of ${\rm BR}(B_{d,s}^0 \to J/\psi \sigma/f_0(\to \pi^+ \pi^-))$
around $\phi_f \approx 25^\circ$ with a twofold ambiguity
from Fig.~\ref{fig:fig2}. Frankly speaking,
this twofold ambiguity cannot be resolved in these considered $B_{d,s}^0 \to
J/\psi \sigma(f_0)$ decays because there are no any interferences
between the final states $J/\psi f_q$ and $J/\psi f_s$. That means it
tends to be resolved through the studies of other $B \to M \sigma(f_0)$ decays
with $M$ denoting the open-charmed or light hadrons, once the related
measurements are available with high precision.

Then, within theoretical uncertainties,
the NLO PQCD predictions of ${\rm BR}(B_{d,s}^0 \to J/\psi \sigma(f_0),
\sigma(f_0) \to \pi^+ \pi^-)$ at $\phi_f \approx 25^\circ$ can be read
as follows:
\beq
{\rm BR}(B_d^0 \to J/\psi \sigma, \sigma \to \pi^+ \pi^-)&=&
1.22^{+0.41}_{-0.29}(\omega_B)^{+0.19}_{-0.17}(f_{J/\psi})^{+0.34}_{-0.29}(B_i^q)
^{+0.13}_{-0.21}(a_t)^{+0.12}_{-0.12}({\cal B}_{\sigma})\times 10^{-5}\non
%^{+0.41+0.19+0.33+0.10+0.13+0.12}_{-0.29-0.17-0.28-0.09-0.21-0.12}
&=&1.22^{+0.60}_{-0.51}\times 10^{-5}\;,
\label{eq:brdpsis0-pipi}\\
{\rm BR}(B_s^0 \to J/\psi f_0, f_0 \to \pi^+ \pi^-)&=&
1.30^{+0.50}_{-0.33}(\omega_B)^{+0.21}_{-0.18}(f_{J/\psi})^{+0.30}_{-0.27}(B_i^s)
^{+0.19}_{-0.23}(a_t)^{+0.20}_{-0.14}({\cal B}_{f_0})
%^{+0.50+0.21+0.29+0.09+0.19+0.20}_{-0.33-0.18-0.25-0.09-0.23-0.14}
\times 10^{-4}\non
&=&1.30^{+0.68}_{-0.53}\times 10^{-4}\;;
\label{eq:brspsif0-pipi}
\eeq
\beq
{\rm BR}(B_d^0 \to J/\psi f_0, f_0 \to \pi^+ \pi^-)&=&
1.64^{+0.54}_{-0.39}(\omega_B)^{+0.26}_{-0.23}(f_{J/\psi})^{+0.46}_{-0.39}(B_i^q)
^{+0.17}_{-0.28}(a_t)^{+0.25}_{-0.18}({\cal B}_{f_0})
%^{+0.54+0.26+0.44+0.14+0.17+0.25}_{-0.39-0.23-0.37-0.13-0.28-0.18}
\times 10^{-6}\non
&=&1.64^{+0.81}_{-0.69}\times 10^{-6}\;,
\label{eq:brdpsif0-pipi}\\
{\rm BR}(B_s^0 \to J/\psi \sigma, \sigma \to \pi^+ \pi^-)&=&
4.56^{+1.74}_{-1.16}(\omega_B)^{+0.71}_{-0.63}(f_{J/\psi})^{+1.06}_{-0.93}(B_i^s)
^{+0.66}_{-0.80}(a_t)^{+0.46}_{-0.45}({\cal B}_{\sigma})
%^{+1.74+0.71+1.01+0.32+0.66+0.46}_{-1.16-0.63-0.88-0.31-0.80-0.45}
\times 10^{-5}\non
&=&4.56^{+2.30}_{-1.86}\times 10^{-5}\;.
\label{eq:brspsis0-pipi}
\eeq
The dominant errors are induced by the shape parameter
$\omega_B = 0.40 \pm 0.04 (\omega_B = 0.50 \pm 0.05)$~GeV for the $B_d^0(B_s^0)$ meson,
the decay constant $f_{J/\psi}= 0.405 \pm 0.014$~GeV for the $J/\psi$ meson,
the Gegenbauer moments $B_{i}^{q,s}$[see Eq.~(\ref{eq:gm-s})] in the leading-twist
light-cone distribution amplitude of light scalar $f_{q,s}$ states,
and the branching ratios ${\cal B}_{\sigma/f_0 \to \pi^+ \pi^-}$, respectively.
Furthermore, we also investigate the higher order contributions simply through
exploring the variation of the running hard scale $t_{\rm max}$, i.e., from $0.8t$ to $1.2t$ (not changing $1/b_i, i= 1,2,3$), in the hard kernel,
which has been counted into one of the sources of theoretical uncertainties.
In every second line of the above equations, various errors have been added
in quadrature.

It is worthwhile to stress that, within still large uncertainties, the NLO PQCD
predictions about the $B_{d}^0 \to J/\psi \sigma(\to \pi^+ \pi^-)$ and
$B_{d,s}^0 \to J/\psi f_0(\to \pi^+ \pi^-)$ decay rates are generally
consistent with the current data or upper limits, except for the seemingly
challenging $B_s^0 \to J/\psi \sigma(\to \pi^+ \pi^-)$ one.
Nevertheless, roughly speaking, the theoretical prediction of
${\rm BR}(B_s^0 \to J/\psi \sigma(\to \pi^+ \pi^-))$ could agree with the current
upper limits within $3\sigma$(not to be confused with the $\sigma$ meson) standard
deviations. Of course, more relevant studies are demanded theoretically and
experimentally.

In order to find more evidences for the consistency between theory and experiment
under the assumption of $\sigma-f_0$ mixing in the conventional two-quark
structure, it is better for us to study
the relative ratios of the above-mentioned branching ratios over
those of the referenced channels such as the
preferred ${\rm BR}(B_s^0 \to J/\psi \phi(\to K^+ K^-))$, because the effects
induced by the uncertainties of nonperturbative inputs are expected to
be canceled to a great extent. This cancellation can also be easily observed
in the quantities such as {\it CP}-violating asymmetries that are
%will be
clarified later. Therefore, following
Eqs.~(\ref{eq:rf0phi-def})-(\ref{eq:R-f02pphi-pdg}), the relative
ratio $R_{f_0/\phi}^{\rm Th.}(\pi)$ in the PQCD approach at NLO accuracy
could be easily obtained as
\beq
R_{f_0/\phi}^{\rm Th.}(\pi) &\equiv&
\frac{{\rm BR}(B_s^0 \to J/\psi f_0) {\cal B}(f_0 \to \pi^+ \pi^-)}
{{\rm BR}(B_s^0 \to J/\psi \phi) {\cal B}(\phi \to K^+ K^-)}\Bigg{|}_{\rm PQCD}
= 0.258^{+0.032}_{-0.041}\;,
\label{eq:R-f02pphi2k-th}
\eeq
and
\beq
\frac{{\rm BR}(B_s^0 \to J/\psi f_0, f_0 \to \pi^+ \pi^-)}
{{\rm BR}(B_s^0 \to J/\psi \phi)}\Bigg{|}_{\rm PQCD}&=& 0.126^{+0.017}_{-0.020}\;,
\label{eq:R-f02pphi-th}
\eeq
assisted with the available values
${\rm BR}(B_s^0 \to J/\psi \phi)|_{\rm PQCD} =
1.02^{+0.36}_{-0.30} \times  10^{-3} $~\cite{Liu:2013nea}
and ${\cal B}(\phi \to K^+ K^-) = 0.492 \pm 0.005$~\cite{Tanabashi:2018oca}.
These two ratios are found to agree well with the measurements as shown in
Eqs.~(\ref{eq:rf0phi-data}) and~(\ref{eq:R-f02pphi-pdg}).

Furthermore, as reported by the LHCb Collaboration, the latest values of
${\rm BR}(B_d^0 \to J/\psi \rho^0, \rho^0 \to \pi^+ \pi^-)$ and ${\rm BR}(B_d^0
\to J/\psi \sigma, \sigma \to \pi^+ \pi^-)$ are as follows~\cite{Aaij:2014siy},
\beq
{\rm BR}(B_d^0 \to J/\psi \rho^0, \rho^0 \to \pi^+ \pi^-)
&=& 2.50^{+0.21}_{-0.18} \times 10^{-5}\;,\label{eq:brpsir02p-ex} \\
{\rm BR}(B_d^0 \to J/\psi \sigma, \sigma \to \pi^+ \pi^-)
&=& 0.88^{+0.12}_{-0.16} \times 10^{-5}\;.\label{eq:brpsis02p-ex}
\eeq
Then the relative ratio of these two branching ratios could be derived
analogously as
\beq
R_{\sigma/\rho}&\equiv&
\frac{{\rm BR}(B_d^0 \to J/\psi \sigma, \sigma \to \pi^+ \pi^-)}
{{\rm BR}(B_d^0 \to J/\psi \rho^0, \rho^0 \to \pi^+ \pi^-)}
= 0.352^{+0.017}_{-0.042}\;,
\label{eq:R-r02ps02p-ex}
\eeq
It is commented that, based on the isospin conservation in the strong interactions,
the branching ratio of $\rho^0 \to \pi^+ \pi^-$ is about
$100\%$~\cite{Tanabashi:2018oca}. Therefore, by combining with the
available prediction ${\rm BR}(B_d^0 \to J/\psi \rho^0)|_{\rm PQCD} =
2.7^{+1.0}_{-0.7} \times 10^{-5}$~\cite{Liu:2013nea} and Eq.~(\ref{eq:brdpsis0-pipi}),
the corresponding ratio predicted theoretically in the PQCD approach
can be read as
\beq
R_{\sigma/\rho}^{\rm Th.}(\pi) &\equiv& \frac{{\rm BR}(B_d^0 \to J/\psi \sigma)
{\cal B}(\sigma \to \pi^+ \pi^-)}{{\rm BR}(B_d^0 \to J/\psi \rho^0)
{\cal B}(\rho^0 \to \pi^+ \pi^-)}\Bigg{|}_{\rm PQCD}
= 0.452^{+0.040}_{-0.097}\;,
\label{eq:R-s02pr02p-th}
\eeq
which is basically consistent with that, see Eq.~(\ref{eq:R-r02ps02p-ex}),
extracted from the LHCb measurement within large errors.
It is clearly observed
that the PQCD predicted branching ratios and the relevant
ratios of $B_{d(s)}^0 \to J/\psi \sigma(f_0)
(\to \pi^+ \pi^-)$ decays with the mixing angle $\phi_f$ around
$25^\circ$ indeed agree with the corresponding measurements within
uncertainties. It is interesting to note that these predictions
are also consistent with those already presented in the
literature~\cite{Fleischer:2011au,Wang:2015uea}.

Similarly, the ratios $R_{f_0/\rho}^{\rm Th.}(\pi)$ and
$R_{\sigma/\phi}^{\rm Th.}(\pi)$ in the PQCD approach could be predicted as
\beq
R_{f_0/\rho}^{\rm Th.}(\pi) &\equiv& \frac{{\rm BR}(B_d^0 \to J/\psi f_0)
{\cal B}(f_0 \to \pi^+ \pi^-)}{{\rm BR}(B_d^0 \to J/\psi \rho^0)
{\cal B}(\rho^0 \to \pi^+ \pi^-)}\Bigg{|}_{\rm PQCD}
= 0.061^{+0.005}_{-0.013}\;,
\label{eq:R-f02pr02p-th} \\
R_{\sigma/\phi}^{\rm Th.}(\pi) &\equiv& \frac{{\rm BR}(B_s^0 \to J/\psi \sigma)
{\cal B}(\sigma \to \pi^+ \pi^-)}{{\rm BR}(B_s^0 \to J/\psi \phi)
{\cal B}(\phi \to K^+ K^-)}\Bigg{|}_{\rm PQCD}
= 0.090^{+0.010}_{-0.014}\;,
\label{eq:R-s02pphi2k-th}
\eeq
which are expected to be examined in the future measurements,
even if the $B_s^0 \to J/\psi \sigma(\sigma \to \pi^+ \pi^-)$ decay
rate highly supersedes the current upper limit set by the
LHCb Collaboration.

From the above results, one can see that most of our PQCD predictions on {\it CP}-averaged branching ratios and relevantly relative ratios of
$B_{d,s}^0 \to J/\psi \sigma/f_0(\to \pi^+ \pi^-)$ up to NLO precision agree well
with the existing experimental measurements within uncertainties at $\phi_f$ around $25^\circ$.
Therefore, the
branching ratios of the decays $B_{d,s}^0 \to J/\psi \sigma(f_0)$ under consideration
in the PQCD approach are presented within errors as follows,
\begin{itemize}
\item {for $\bar b \to \bar d$ decay channels,}
\beq
{\rm BR}(B_d^0 \to J/\psi \sigma) &=&
1.83^{+0.61}_{-0.43}(\omega_B)
%^{+0.2+0.1+0.4}_{-0.1-0.0-0.3}(f_M)
^{+0.29}_{-0.25}(f_{J/\psi})
%^{+0.1+0.2}_{-0.1-0.1}(a_i)
^{+0.51}_{-0.44}(B_i^q)
%^{+0.49+0.15}_{-0.42-0.14}(B_i^q)
%^{+1.03}_{-0.}(m_c)
^{+0.19}_{-0.31}(a_t)
[1.83^{+0.87}_{-0.73}]
%^{+0.0}_{-0.0}({\rm CKM})
\times  10^{-5}  \label{eq:brdpsis0} , \\
{\rm BR}(B_d^0 \to J/\psi f_0) &=&
3.64^{+1.21}_{-0.86}(\omega_{B})
^{+0.57}_{-0.51}(f_{J/\psi})
^{+1.02}_{-0.88}(B_{i}^q)
%^{+0.97+0.30}_{-0.83-0.28}(B_{i}^q)
%^{+2.04}_{-0.9}(m_c)
^{+0.37}_{-0.62}(a_t)
[3.64^{+1.72}_{-1.47}]
%^{+0.0}_{-0.0}({\rm CKM})
\times  10^{-6}  \label{eq:brdpsif0} ;
\eeq

\item {for $\bar b \to \bar s$ decay channels,}
\beq
{\rm BR}(B_s^0 \to J/\psi \sigma)  &=&
6.83^{+2.61}_{-1.74}(\omega_{B})
%^{+0.15+0.03+0.06}_{-0.14-0.03-0.06}(f_M)
^{+1.07}_{-0.95}(f_{J/\psi})
%^{+0.02+0.07}_{-0.02-0.06}(a_{i})
^{+1.58}_{-1.40}(B_{i}^s)
%^{+1.51+0.48}_{-1.32-0.46}(B_{i}^s)
%^{+0.22}_{-0.20}(m_c)
^{+0.99}_{-1.20}(a_t)
[6.83^{+3.38}_{-2.71}]
%^{+0.0}_{-0.0}({\rm CKM})
\times  10^{-5}  \label{eq:brspsis0} ,\\
{\rm BR}(B_s^0 \to J/\psi f_0) &=&
2.89^{+1.11}_{-0.73}(\omega_B)
^{+0.46}_{-0.40}(f_{J/\psi})
^{+0.67}_{-0.58}(B_{i}^s)
%^{+0.64+0.21}_{-0.55-0.19}(B_{i}^s)
%^{+0.24}_{-0.22}(m_c)
^{+0.42}_{-0.50}(a_t)
[2.89^{+1.44}_{-1.13}]
%^{+0.0}_{-0.0}({\rm CKM})
\times  10^{-4}  \label{eq:brspsif0}  ,
\eeq
\end{itemize}
where, as shown in the square brackets, various errors of the
numerical results have also been added in quadrature.
One can observe that the decay rates
for the $\bar b \to \bar s$ transition processes, i.e.,
$B_{s}^0 \to J/\psi \sigma(f_0)$,
are generally much larger than those for the $\bar b \to \bar d$ transition ones,
i.e., $B_{d}^0 \to J/\psi \sigma(f_0)$.
This is due to the CKM hierarchy for two kinds of processes:
the CKM factors $V_{cb}V_{cs}$  in $b\to s$  are about four times larger
than the  $V_{cb}V_{cd}$  for $b\to d$ process, and the different
factors $\sin^2\phi_f$ or $\cos^2\phi_f$ from the mixtures of $\sigma$
and $f_0$ mesons.  The remanent but small differences arise from the SU(3)
symmetry breaking effects in the hadronic parameters, such as
decay constants, mesonic masses, distribution amplitudes, etc..
It is easily seen that our NLO PQCD predicted branching ratios of the
$B_{d,s}^0 \to J/\psi \sigma(f_0)$ decays around $\phi_f \approx 25^\circ$
are generally consistent with those earlier predictions~\cite{Colangelo:2010bg,Colangelo:2010wg,Leitner:2010fq,Fleischer:2011au,
Li:2012sw} as aforementioned in the introduction within still large
uncertainties.

Based on those PQCD branching ratios as presented in the
Eqs.~(\ref{eq:brdpsis0})-(\ref{eq:brspsif0}),
several interesting ratios could be derived as follows:
\beq
R_{\sigma f_0}^{d}&\equiv& \frac{{\rm BR}(B_d^0 \to J/\psi \sigma)}
{{\rm BR}(B_d^0 \to J/\psi f_0)}\Bigg{|}_{\rm PQCD}
(\approx 5.03^{+0.02}_{-0.01})=
\frac{\Phi_\sigma^d}{\Phi^d_{f_0}}\cdot \cot^2\phi_f
%\non
%&\approx& 5.03^{+0.02}_{-0.01}
%^{+0.01+0.02+0.00+0.00+0.01}_{-0.00-0.00-0.01-0.00-0.00}
\;,
\label{eq:R-sf-d} \\
R_{f_0\sigma}^s &\equiv& \frac{{\rm BR}(B_s^0 \to J/\psi f_0)}
{{\rm BR}(B_s^0 \to J/\psi \sigma)}\Bigg{|}_{\rm PQCD}
(\approx 4.23^{+0.03}_{-0.00})
= \frac{\Phi_{f_0}^s}{\Phi_\sigma^s}\cdot \cot^2\phi_f
%\non
%&\approx& 4.23^{+0.03}_{-0.00}
%^{+0.01+0.01+0.02+0.01}_{-0.00-0.00-0.00-0.00} \;;
\label{eq:R-sf-s}
\eeq
\beq
R^{\sigma}_{sd} &\equiv& \frac{{\rm BR}(B_s^0 \to J/\psi \sigma)}
{{\rm BR}(B_d^0 \to J/\psi \sigma)}\Bigg{|}_{\rm PQCD}
(\approx 3.73^{+0.27}_{-0.17})
\non
&=&
\frac{\tau_{B_s^0}}{\tau_{B_d^0}}\cdot (\frac{m_{B_s^0}}{m_{B_d^0}})^7
\cdot \frac{\Phi_\sigma^s}{\Phi_\sigma^d} \cdot
\frac{|{\cal A}(B_s^0 \to J/\psi f_s)/m_{B_s^0}^2|^2}
{|{\cal A}(B_d^0 \to J/\psi f_q)/m_{B_d^0}^2|^2} \cdot \tan^2\phi_f
%^{+0.15+0.00+0.17+0.15}_{-0.10-0.01-0.13-0.03}
\;,
\label{eq:R-sd-s0}\\
R^{f_0}_{sd} &\equiv& \frac{{\rm BR}(B_s^0 \to J/\psi f_0)}
{{\rm BR}(B_d^0 \to J/\psi f_0)}\Bigg{|}_{\rm PQCD}
(\approx 79.43^{+6.13}_{-3.45})
\non
&=&
\frac{\tau_{B_s^0}}{\tau_{B_d^0}}\cdot (\frac{m_{B_s^0}}{m_{B_d^0}})^7
\cdot \frac{\Phi_{f_0}^s}{\Phi_{f_0}^d} \cdot
\frac{|{\cal A}(B_s^0 \to J/\psi f_s)/m_{B_s^0}^2|^2}
{|{\cal A}(B_d^0 \to J/\psi f_q)/m_{B_d^0}^2|^2} \cdot \cot^2\phi_f
%^{+3.21+0.25+4.12+3.19}_{-1.87-0.06-2.88-0.37}
\;.
\label{eq:R-sd-f0}
\eeq
\beq
R_{d \sigma}^{s f_0} &\equiv& \frac{{\rm BR}(B_s^0 \to J/\psi f_0)}
{{\rm BR}(B_d^0 \to J/\psi \sigma)}\Bigg{|}_{\rm PQCD}
(\approx 15.8^{+1.2}_{-0.7})
\non
&=&
\frac{\tau_{B_s^0}}{\tau_{B_d^0}}\cdot (\frac{m_{B_s^0}}{m_{B_d^0}})^7
\cdot \frac{\Phi_{f_0}^s}{\Phi_{\sigma}^d} \cdot
\frac{|{\cal A}(B_s^0 \to J/\psi f_s)/m_{B_s^0}^2|^2}
{|{\cal A}(B_d^0 \to J/\psi f_q)/m_{B_d^0}^2|^2}
%^{+0.6+0.0+0.8+0.6}_{-0.4-0.1-0.6-0.1} \;,
\label{eq:R-sf0-ds0}\\
R_{d f_0}^{s \sigma} &\equiv& \frac{{\rm BR}(B_s^0 \to J/\psi \sigma)}
{{\rm BR}(B_d^0 \to J/\psi f_0)}\Bigg{|}_{\rm PQCD}
(\approx 18.7^{+1.4}_{-0.9})
\non
&=&
\frac{\tau_{B_s^0}}{\tau_{B_d^0}}\cdot (\frac{m_{B_s^0}}{m_{B_d^0}})^7
\cdot \frac{\Phi_{\sigma}^s}{\Phi_{f_0}^d} \cdot
\frac{|{\cal A}(B_s^0 \to J/\psi f_s)/m_{B_s^0}^2|^2}
{|{\cal A}(B_d^0 \to J/\psi f_q)/m_{B_d^0}^2|^2}
%^{+0.7+0.0+0.9+0.8}_{-0.5-0.0-0.7-0.1}\;.
\label{eq:R-ss0-df0}
\eeq
Then, some remarks are in order.
\begin{itemize}
\item[(a)]
It is interesting to note that the first two ratios $R_{\sigma f_0}^d$
and $R^s_{f_0\sigma}$ in the PQCD approach are almost invariant to the aforementioned
various nonperturbative parameters, although the corresponding branching
ratios show strong sensitivity to them. Again, the effects induced by various errors in the relevant branching ratios have been canceled significantly. Thus, as discussed in the
literature, e.g., Refs.~\cite{Stone:2013eaa} and~\cite{Li:2012sw}, these
two relations could be utilized to extract the angle $\phi_f$ between
$\sigma$ and $f_0$ mixing in the two-quark picture cleanly,
because $R_{\sigma f_0}^d$ and $R^s_{f_0\sigma}$ are almost
equal to $\cot^2\phi_f$ with the almost definite values
$\Phi_\sigma^d/\Phi_{f_0}^d \approx 1.095$
and $\Phi_\sigma^s/\Phi_{f_0}^s \approx 1.087$, respectively.

\item[(b)]
 As presented in the last two ratios, $R_{d \sigma}^{s f_0}$
and $R_{d f_0}^{s \sigma}$ are independent on the mixing angle $\phi_f$, and
%which
are of great interest to examine the SU(3) flavor symmetry breaking
effects, if the penguin contributions are indeed tiny and negligible. To see
more explicitly, these two ratios could be further derived by factoring out
the related CKM matrix elements $V_{cs}$ and $V_{cd}$ correspondingly,
\beq
R_{d \sigma}^{s f_0}&=& \frac{\tau_{B_s^0}}{\tau_{B_d^0}}\cdot
(\frac{m_{B_s^0}}{m_{B_d^0}})^7 \cdot \frac{\Phi_{f_0}^s}{\Phi_{\sigma}^d}
\cdot \frac{|V_{cs}|^2}{|V_{cd}|^2} \cdot
\frac{|{\cal A}^\prime(B_s^0 \to J/\psi f_s) |^2}
{|{\cal A}^\prime(B_d^0 \to J/\psi f_q) |^2}\;,
\label{eq:Rp-sf0-ds0}\\
R_{d f_0}^{s \sigma}&=& \frac{\tau_{B_s^0}}{\tau_{B_d^0}}\cdot
(\frac{m_{B_s^0}}{m_{B_d^0}})^7 \cdot \frac{\Phi_{\sigma}^s}{\Phi_{f_0}^d}
\cdot \frac{|V_{cs}|^2}{|V_{cd}|^2} \cdot
\frac{|{\cal A}^\prime(B_s^0 \to J/\psi f_s)|^2}
{|{\cal A}^\prime(B_d^0 \to J/\psi f_q)|^2} \;,
\label{eq:Rp-ss0-df0}
\eeq
which consequently result in $\frac{|{\cal A}^\prime(B_s^0 \to J/\psi
f_s)|^2}{|{\cal A}^\prime(B_d^0 \to J/\psi f_q)|^2} \approx 0.72$, deviating
from unity about 30\% roughly. Here, ${\cal A}^\prime \equiv {\cal A}/m_{B}^2$.

\item[(c)]
In light of the above-mentioned two points, it seems more complicated that
the entanglement of the SU(3) symmetry breaking effects and the information of
mixing angle $\phi_f$ exhibits evidently in the middle two relations.
Nevertheless,
these two ratios could provide constraints supplementarily to
either the former or the latter when one of them in the first two or
last two ratios could be manifested definitely.
\end{itemize}

By the way, the mixing angle $\phi_f$ can also be constrained similarly
from the ratios of the measured $B_d^0 \to J/\psi \sigma$ and $B_s^0 \to
J/\psi f_0$ decays over the referenced $B_d^0 \to J/\psi \rho^0$ and
$B_s^0 \to J/\psi \phi$ ones with high precision, respectively,
but suffer probably from nonperturbative pollution
induced by the hadronic parameters.

%%%%%%%%%%%%%%%%%%%%%%%%%%%%%%%%%%%%%%%%%%%%%%%%%%%%%%%%%%
%\subsection{CP-violating Asymmetries}\label{ssec:cp-cpas}
%%%%%%%%%%%%%%%%%%%%%%%%%%%%%%%%%%%%%%%%%%%%%%%%%%%%%%%%%%

Now, let us turn to analyze
the {\it CP} violations of the $B_{d,s}^0 \to J/\psi \sigma(f_0)$ decays
in the PQCD approach at NLO accuracy.
As for the {\it CP}-violating asymmetries for the $B_{d,s}^0 \to
J/\psi \sigma(f_0)$ decays, the effects of neutral $B_{d,s}^0-\bar{B}_{d,s}^0$
mixing should be taken into account. The {\it CP}-violating asymmetries
of $B_{d,s}^0(\bar{B}_{d,s}^0) \to J/\psi \sigma(f_0)$ decays are time
dependent and can be defined as
\beq
A_{\rm CP} &\equiv& \frac{\Gamma\left
(\bar{B}_{d,s}^0(\Delta t) \to f_{\rm CP}\right) -
\Gamma\left(B_{d,s}^0(\Delta t) \to f_{\rm CP}\right )}{ \Gamma\left
(\bar{B}_{d,s}^0(\Delta t) \to f_{\rm CP}\right ) + \Gamma\left
(B_{d,s}^0(\Delta t) \to f_{\rm CP}\right ) }\non
&=& A_{\rm CP}^{\rm dir} \cos(\Delta m_{d,s}  \Delta t)
+ A_{\rm CP}^{\rm mix} \sin (\Delta m_{d,s} \Delta
t)\;, \label{eq:acp-def}
\eeq
where $\Delta m_{d,s}$ is the mass
difference between the two $B_{d,s}^0$ mass eigenstates, $\Delta
t =t_{\rm CP}-t_{tag} $ is the time difference between the tagged
$B_{d,s}^0$ ($\bar{B}_{d,s}^0$) and the accompanying
$\bar{B}_{d,s}^0$ ($B_{d,s}^0$) with opposite $b$ flavor
decaying to the final {\it CP} eigenstate $f_{\rm CP}$ at the time $t_{\rm CP}$.
The direct and mixing-induced {\it CP}-violating asymmetries
$A_{\rm CP}^{\rm dir}({\cal C}_f)$ and $A_{\rm CP}^{\rm mix}({\cal S}_f)$
can be written as
\beq
A_{\rm CP}^{\rm dir}&\equiv& {\cal C}_f = \frac{ \left |
\lambda_{\rm CP}^{d,s}\right |^2-1 } {1+\left |\lambda_{\rm CP}^{d,s}\right |^2},
\qquad
A_{\rm CP}^{\rm mix}\equiv{\cal S}_f= \frac{ 2 {\rm Im}
(\lambda_{\rm CP}^{d,s})}{1+\left |\lambda_{\rm CP}^{d,s}\right |^2},
\label{eq:acp-csf}
\eeq
with the {\it CP}-violating parameter $\lambda_{\rm CP}^{d,s}$,
\beq
\lambda_{\rm CP}^{d,s} &\equiv& \eta_f \; \frac{V_{tb}^*V_{td(s)}}{V_{tb}V_{td(s)}^*}
\cdot \frac{ \langle f_{\rm CP} |H_{\rm eff}|\bar{B}_{d,s}^0\rangle}
{\langle f_{\rm CP} |H_{\rm eff}|B_{d,s}^0\rangle},
\label{eq:lambda}
\eeq
where $\eta_f$ is the {\it CP} eigenvalue of the final states.
Moreover, for $B_s^0$ meson decays, a nonzero ratio $(\Delta
\Gamma/\Gamma)_{B_s^0}$ is expected in the
SM~\cite{Beneke:1998sy,Fernandez:2006qx}.
For $B_s^0 \to J/\psi \sigma(f_0)$ decays, the third term
$A_{\rm CP}^{\Delta \Gamma_s}$ related
to the presence of a non-negligible $\Delta \Gamma_s$
to describe the {\it CP} violation can be defined as
follows~\cite{Fernandez:2006qx}:
\beq
A_{\rm CP}^{\Delta \Gamma_s} &=& \frac{ 2 {\rm Re}
( \lambda_{\rm CP}^{s})}{1+\left |\lambda_{\rm CP}^{s}\right |^2}.
\label{eq:acp-dgs}
\eeq
The above three quantities describing the {\it CP} violations in $B_s^0$
meson decays shown in Eqs.~(\ref{eq:acp-csf}) and
(\ref{eq:acp-dgs}) satisfy the following relation,
\beq
|A_{\rm CP}^{\rm dir}|^2+ |A_{\rm CP}^{\rm mix}|^2
 + |A_{\rm CP}^{\Delta \Gamma_s}|^2 &=&
 1 \;.\label{eq:summation-cp}
\eeq

The {\it CP}-violating parameters $\lambda_{\rm CP}^{d}$ and $\lambda_{\rm CP}^{s}$
defined for the $B_d^0 \to J/\psi \sigma(f_0)$ and $B_s^0 \to J/\psi \sigma(f_0)$
decays can be written explicitly as
\beq
\lambda_{\rm CP}^d &=& \eta_{f}\frac{V_{tb}^* V_{td}}{V_{tb} V_{td}^*}
\cdot \frac{\overline{\cal A}(\bar B_d^0 \to J/\psi \sigma(f_0))}
{{\cal A}(B_d^0 \to J/\psi \sigma(f_0))}\;,\qquad
\lambda_{\rm CP}^s = \eta_{f}\frac{V_{tb}^* V_{ts}}{V_{tb} V_{ts}^*}
\cdot \frac{\overline{\cal A}(\bar B_s^0 \to J/\psi \sigma(f_0))}
{{\cal A}(B_s^0 \to J/\psi \sigma(f_0))}\;,
\eeq
with the {\it CP} eigenvalue $\eta_{f} = -1$.
Based on Eqs.~(\ref{eq:decay-dps0})-(\ref{eq:decay-spf0}), it is easy to
observe that $\lambda_{\rm CP}^d$
and $\lambda_{\rm CP}^s$ are actually determined by the decay
amplitudes of $B_d^0 \to J/\psi f_q$ and $B_s^0 \to J/\psi f_s$, respectively.
The results of $\lambda_{\rm CP}^d$
and $\lambda_{\rm CP}^s$ can then be read numerically as
\beq
\lambda_{\rm CP}^d &=& (-0.709^{+0.000}_{-0.001})
+ {\it i} (0.681^{+0.000}_{-0.001})
%(-0.709^{+0.000}_{-0.001-0.001}) + {\it i} (0.681^{+0.000}_{-0.000-0.001})
\;,\label{eq:lambda-d} \\
\lambda_{\rm CP}^s &=& (-1.000^{+0.000}_{-0.000})
- {\it i} (0.037^{+0.000}_{-0.000})\;.
\label{eq:lambda-s}
\eeq
Therefore, their modules can be read correspondingly as,
\beq
|\lambda_{\rm CP}^d| &=& 0.983^{+0.001}_{-0.000}\;,\\
|\lambda_{\rm CP}^s| &=& 1.001^{+0.000}_{-0.000}\;,
\eeq
which indicate a slightly large(tiny) penguin contamination in these
considered $B_d^0(B_s^0)$ decay modes. It is interesting to note
that the consistent measurement of $|\lambda|=1.01^{+0.08}_{-0.06}
\pm 0.03$(the first uncertainty is statical and the second systematic)
in the $B_s^0 \to J/\psi \pi^+ \pi^-$ decay
was reported very recently by the LHCb Collaboration~\cite{Aaij:2019mhf}.

Then, the {\it CP} violations of
$B_{d,s}^0 \to J/\psi \sigma(f_0)$ in the PQCD approach
are as follows,
\beq
A_{\rm CP}^{\rm dir}(B_d^0 \to J/\psi \sigma(f_0))
&\equiv&
A_{\rm CP}^{\rm dir}(B_d^0 \to J/\psi f_q)
= -1.70^{+0.06}_{-0.06}
%^{+0.05+0.03}_{-0.05-0.03}
\times 10^{-2}\;, \\
A_{\rm CP}^{\rm mix}(B_d^0 \to J/\psi \sigma(f_0))
&\equiv&
A_{\rm CP}^{\rm mix}(B_d^0 \to J/\psi f_q)
= 0.692^{+0.001}_{-0.000}
%^{+0.001+0.001}_{-0.000-0.000}
\;;\label{eq:cpm-d}
\eeq
\beq
A_{\rm CP}^{\rm dir}(B_s^0 \to J/\psi \sigma(f_0))
&\equiv&
A_{\rm CP}^{\rm dir}(B_s^0 \to J/\psi f_s)
= 0.733^{+0.032}_{-0.044}
%^{+0.028+0.016}_{-0.041-0.016}
\times 10^{-3}\;,\\
A_{\rm CP}^{\rm mix}(B_s^0 \to J/\psi \sigma(f_0))
&\equiv&
A_{\rm CP}^{\rm mix}(B_s^0 \to J/\psi f_s)
= -3.70^{+0.00}_{-0.01} \times 10^{-2}\;,
\label{eq:cpm-s}\\
A_{\rm CP}^{\rm \Delta\Gamma_s}(B_s^0 \to J/\psi \sigma(f_0))
&\equiv&
A_{\rm CP}^{\rm \Delta\Gamma_s}(B_s^0 \to J/\psi f_s)
= -0.999^{+0.000}_{-0.000}\;.
\eeq
Notice that a {\it CP}-violating effect
$\alpha_{\rm CP} = \frac{1-|\lambda_f|}{1+|\lambda_f|}$ with
$\lambda_f$ being the {\it CP}-violating parameter like $\lambda_{\rm CP}^d$
is fitted as $-58 \pm 46 \times 10^{-3}$ for resonance $f_0(500)$ in the $B^0 \to J/\psi \pi^+ \pi^-$ decays~\cite{Aaij:2014vda}, which is roughly
consistent with our prediction within still large experimental errors.

The above two mixing-induced {\it CP} violations,
i.e., Eqs.~(\ref{eq:cpm-d}) and~(\ref{eq:cpm-s}), could
be utilized to estimate the penguin
impacts on the weak
phase $\phi_{d,s}$ in the $B_{d,s}^0 \to J/\psi
\sigma(f_0)$ decays,
\beq
 \phi_{d(s)}^{\rm eff} &=&-  {\rm arg} \left[\Biggl(\frac{q}{p}\Biggr)_{d(s)}
 \frac{\overline {\cal A}^{d(s)}_{f}}{ {\cal A}^{d(s)}_{f}} \right]
 = \phi_{d(s)}^{\rm SM} +\Delta \phi_{d(s)}\;,
\eeq
where ${\cal A}^{d(s)}_f$ and $\overline {\cal A}^{d(s)}_f$ are the decay
amplitudes of $B_{d(s)}^0 \to J/\psi \sigma(f_0)$ and $\bar B_{d(s)}^0 \to
J/\psi \sigma(f_0)$ decays, respectively. In light of the above-mentioned
slightly small or tiny penguin pollution in the $B_{d,s}^0 \to J/\psi \sigma(f_0)$
modes, the mixing-induced {\it CP}-violating asymmetries could be further
written approximately as $A_{\rm CP}^{\rm mix} \equiv {\cal S}_f \simeq
\sin\phi^{\rm eff}$, whose evidently nonzero deviations to the SM one
$\sin\phi^{\rm SM}$ would be helpful to justify the new physics signals beyond SM.
It is worth
%y of
pointing out that only the perturbative expansions at NLO
in $\alpha_s$ and at leading power in $1/m_b$ are taken into account in
the calculations of this work. We extract the quantity $\Delta\phi_s$
from our NLO PQCD evaluations with $t$-quark penguin contributions
as follows,
\beq
\Delta\phi_s &\approx& -0.38^{+0.06}_{-0.04}
% %^{+0.04+0.05}_{-0.03-0.03}
 \times 10^{-3} \;,
\eeq
where the dominant errors are from the variation of the shape parameter
$\omega_{B}$ in the distribution amplitude of $B_s^0$ meson and the Gegenbauer
moments $B_{i}^s$ in the distribution amplitude of flavor state $f_s$,
and various uncertainties have been added in quadrature.
The penguin corrections such as $u$-quark and $c$-quark
loop contributions are not included here.
As discussed in Refs.~\cite{Li:2006vq} and~\cite{Liu:2013nea}, the former
correction demands a two-loop calculation for the corresponding amplitude,
which is not available currently,
%yet
while the latter one does
%will
not contribute
to the quantity $\Delta\phi_s$. Therefore, the more precise value
about $\Delta\phi_s$ extracted from the $B_s^0 \to J/\psi f_0$ mode
by including $u$-quark penguin
contamination has to be presented elsewhere in the future.

Here, we also calculate the modules of amplitudes for the $B_d^0 \to J/\psi \sigma$,
$B_d^0 \to J/\psi f_0$, and $B_s^0 \to J/\psi f_0$ decays with definitions as
$|{\cal A}_d^\sigma|$, $|{\cal A}_d^{f_0}|$, and $|{\cal A}_{s}^{f_0}|$
(in units of GeV$^{3}$),
 \beq
 |{\cal A}_d^\sigma| &\equiv& |{\cal A}(B_d^0 \to J/\psi \sigma)|_{\rm PQCD} \approx
 7.03^{+1.21}_{-1.03}
 %^{+1.09+0.53}_{-0.89-0.51}
 \times 10^{-3}   \;, \\
 |{\cal A}_d^{f_0}| &\equiv& |{\cal A}(B_d^0 \to J/\psi f_0)|_{\rm PQCD} \approx
 3.28^{+0.56}_{-0.48}
 %^{+0.51+0.24}_{-0.41-0.24}
 \times 10^{-3}  \;, \\
 |{\cal A}_s^{f_0}| &\equiv& |{\cal A}(B_s^0 \to J/\psi f_0)|_{\rm PQCD} \approx
 2.89^{+0.56}_{-0.45}
 %^{+0.52+0.22}_{-0.40-0.21}
 \times 10^{-2} \;,
 \eeq
which result in the ratios ${R}_{d/s}^{\sigma f_0}$ between
$|{\cal A}(B_d^0 \to J/\psi \sigma)|$ and $|{\cal A}(B_s^0 \to J/\psi f_0)|$,
and ${R}_{d/s}^{f_0 f_0}$ between $|{\cal A}(B_d^0 \to J/\psi f_0)|$ and
$|{\cal A}(B_s^0 \to J/\psi f_0)|$ as follows,
 \beq
 {R}_{d/s}^{\sigma f_0} &\equiv&
 \Biggl|\frac{{\cal A}_d^\sigma}{{\cal A}_s^{f_0}}\Biggr|_{\rm PQCD}
 = 0.243^{+0.003}_{-0.005}
 %^{+0.003+0.000}_{-0.005-0.000}
 \;,
 \eeq
 \beq
 {R}_{d/s}^{f_0 f_0} &\equiv&
 \Biggl|\frac{{\cal A}_d^{f_0}}{{\cal A}_s^{f_0}}\Biggr|_{\rm PQCD}
 = 0.113^{+0.002}_{-0.002}
 %^{+0.002+0.000}_{-0.002-0.000}
 \;.
 \eeq
These two ratios are expected to be helpful to examine the SU(3) flavor
symmetry breaking effects, as well as the useful information on the
mixing angle $\phi_f$, in these considered $B_{d}^0 \to J/\psi \sigma(f_0)$
and $B_s^0 \to J/\psi f_0$ decays.

%%%%%%%%%%%%%%%%%%%%%%%%%%%%%%%%%%%%%%%%%%%%%%%%%%%%%%%%%%%%%%%%%%%%%%%%%%%%%%%%%%%%%%%%%%%%%%%%%%
\begin{figure}[!!hbt]
\begin{center}
\hspace{-1 cm}
\includegraphics[scale=0.55]{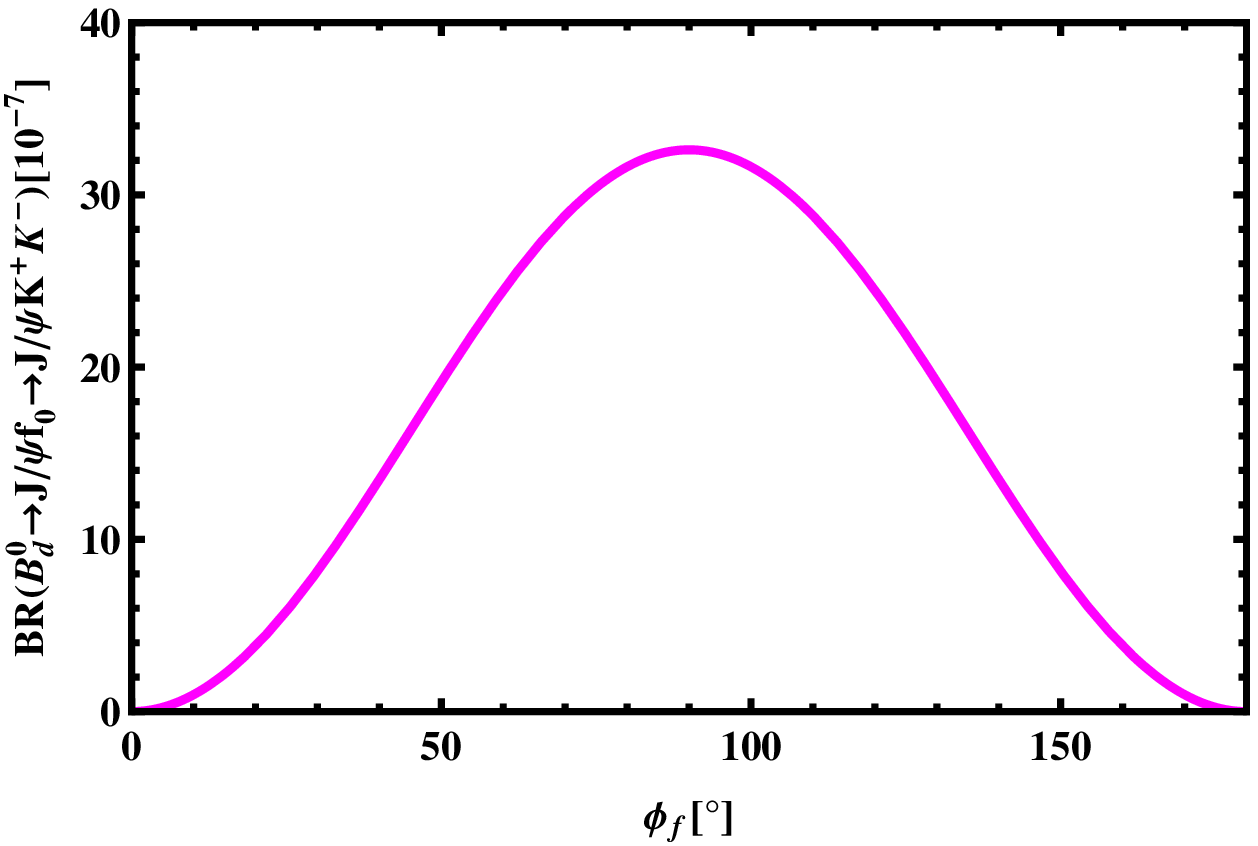}\hspace{0.6cm}
\includegraphics[scale=0.55]{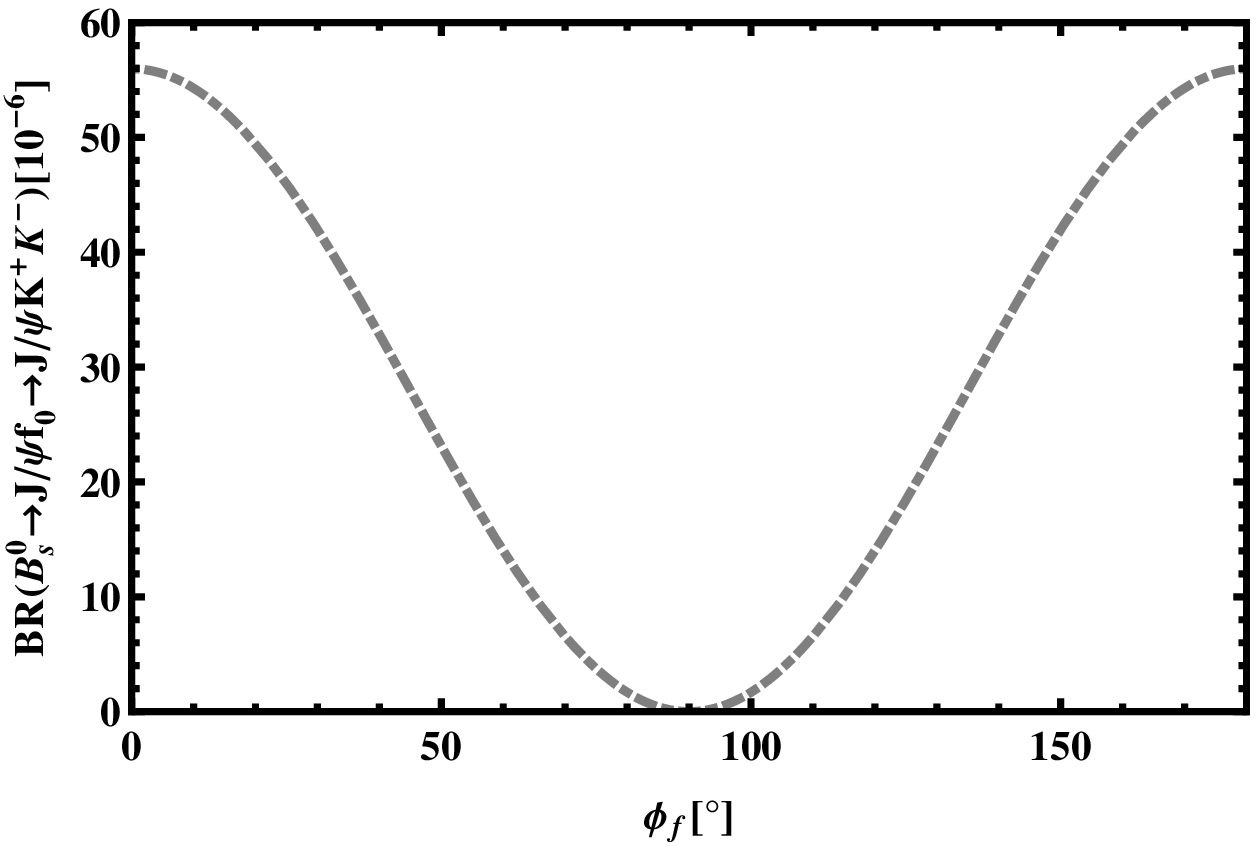}
\vspace{-0.5cm}
\caption{   Dependence on the mixing angle $\phi_{f}$ of the central values
for ${\rm BR}(B_{d,s}^0 \to J/\psi f_0 \to J/\psi K^+ K^-)$ in the PQCD approach: The magenta solid\ [gray dot-dashed] line corresponds to the $B_d^0 \to J/\psi f_0(\to K^+ K^-)\ [B_s^0 \to J/\psi f_0(\to K^+ K^-)]$ decay, respectively.}
\label{fig:fig3}
\end{center}
\end{figure}
%%%%%%%%%%%%%%%%%%%%%%%%%%%%%%%%%%%%%%%%%%%%%%%%%%%%%%%%%%%%%%%%%%%%%%%%%%%%%%%%%%%%%%%%%%%%%%%%%%

Last but not least, it is noted that the scalar meson $f_0$
decays largely into $\pi^+ \pi^-$ but can also
decay into $K^+ K^-$. Therefore, some useful information
about this $f_0$ meson could also be
hinted from the analysis of $B_{d,s}^0 \to J/\psi f_0 \to J/\psi K^+ K^-$ decays.
The dependence of ${\rm BR}(B_{d,s}^0 \to J/\psi f_0(\to K^+ K^-))$ on the mixing
angle $\phi_f$ is plotted in Fig.~\ref{fig:fig3}.
According to ${\cal B}(f_0 \to K^+ K^-)=0.16^{+0.04}_{-0.05}$,
the branching ratios of
$B_{d,s}^0 \to J/\psi f_0(\to K^+ K^-)$, as a byproduct, could be easily obtained
at $\phi_f \approx 25^\circ$ as follows,
\beq
{\rm BR}(B_d^0 \to J/\psi f_0, f_0 \to K^+ K^-)|_{\rm PQCD} &=& 0.58^{+0.31}_{-0.29}
%^{+0.19+0.09+0.16+0.05+0.06+0.15}_{-0.14-0.08-0.13-0.04-0.10-0.18}
\times 10^{-6}  \;,\\
{\rm BR}(B_s^0 \to J/\psi f_0, f_0 \to K^+ K^-)|_{\rm PQCD} &=& 0.46^{+0.26}_{-0.23}
%^{+0.18+0.07+0.10+0.03+0.07+0.11}_{-0.12-0.06-0.09-0.03-0.08-0.14}
\times 10^{-4}  \;.
\eeq
Then, the interesting ratios could be further derived as
\beq
R_{f_0/\rho}^{\rm Th.}(K) &\equiv&
\frac{{\rm BR}(B_d^0 \to J/\psi f_0){\cal B}(f_0 \to K^+ K^-)}
{{\rm BR}(B_d^0 \to J/\psi \rho^0) {\cal B}(\rho^0 \to \pi^+ \pi^-)}\Bigg{|}_{\rm PQCD}
= 0.021^{+0.003}_{-0.006}\;,\\
R_{f_0/\phi}^{\rm Th.}(K) &\equiv&
\frac{{\rm BR}(B_s^0 \to J/\psi f_0) {\cal B} (f_0 \to K^+ K^-)}
{{\rm BR}(B_s^0 \to J/\psi \phi) {\cal B}(\phi \to K^+ K^-)}\Bigg{|}_{\rm PQCD}
= 0.092^{+0.014}_{-0.026}\;.
\eeq
which are expected to be tested in the
measurements at LHCb and/or Belle-II experiments. Furthermore,
the relevant examinations
%will
provide more supplementary
constraints on the mixing angle $\phi_f$.
By the way, frankly speaking, the $B_{s}^0 \to J/\psi f_0(\to K^+ K^-)$
branching ratio measurement is still necessary, although it is very difficult
experimentally as $f_0$ is buried under the tail of $\phi$ (see Fig.~7 in
Ref.~\cite{Aaij:2017zgz} for example)~\cite{Stone:2019ju}.

Finally, two more comments are as follows:
\begin{itemize}
\item [(a)]
For final state interactions:
As mentioned in the above, we just include the short distance contributions
that can be perturbatively calculated in this work. Other possible contributions
such as rescattering effects or final state interactions are not considered
yet, though they are generally believed to affect the predictions of the
observables potentially.

\item [(b)]
For possible tetraquark structure:
In principle, we also need to make some calculations to help identify the
possible tetraquark structure of $\sigma$ and $f_0$. However, the essential
inputs such as light-cone distribution amplitudes are still unavailable now.
Therefore, we cannot obtain the information about the possible tetraquark
components straightforwardly from the perturbative evaluations in the heavy
$B$ meson decays currently.

\end{itemize}
The above two issues have to be left for future investigations after
precise measurements experimentally and related improvements theoretically.

%%%%%%%%%%%%%%%%%%%%%%%%%%%%%%%%%%%%%%
\section{Summary} \label{sec:summary}
%%%%%%%%%%%%%%%%%%%%%%%%%%%%%%%%%%%%%%

As an ideally alternative channel with no need of angular decomposition,
the $B_s^0 \to J/\psi f_0$ decay is expected to have great potential
to reduce errors in the extraction of the $B_s^0-\bar B_s^0$ mixing
phase $\phi_s$, which
will help us to search for the new physics beyond SM associated
with the precision measurements performed at the upgraded LHCb and/or the ongoing
Belle-II experiments. The quantitative exploration demands the reliable
calculations about the corresponding decay amplitude.
As a possible reference, we made the investigations by assuming $f_0$
as the ground scalar meson in the two-quark picture, where it is believed
that $\sigma$ and $f_0$ could mix with each other in the quark-flavor basis with
a single mixing angle $\phi_f$. Up to now, $\phi_f$  has
%is
not been determined
%yet
definitely, although several studies at both theoretical and experimental
aspects have been presented.

Motivated by the global agreement on the observables of the $B \to J/\psi V$
decays between the data and the PQCD approach at NLO accuracy, we extended
that formalism to the $B_{d,s}^0 \to J/\psi \sigma(f_0)$ channels. The NLO PQCD
predictions on the {\it CP}-averaged branching ratios for the $B_{d,s}^0 \to
J/\psi \sigma/f_0 (\to \pi^+ \pi^-)$ decays and the relative ratios generally
agree with the current data or
upper limits within still large
theoretical errors around the mixing angle $\phi_f \approx 25^\circ$ with a twofold
ambiguity. It is stressed that this twofold ambiguity could be resolved
in the $B \to M \sigma(f_0)$ decays with $M$ being certain light or
open-charmed hadrons due to the constructive or destructive interferences
between $B \to M f_q$ and $B \to M f_s$ decays.
Several interesting observables such as branching ratios, relative ratios, and
{\it CP}-violating asymmetries
for the $B_{d,s}^0 \to J/\psi \sigma(f_0)$ decays are then
predicted in the PQCD approach at NLO level. They could be utilized
to either constrain the mixing angle $\phi_f$ or estimate the SU(3)
flavor symmetry breaking effects. As a byproduct, the branching ratios
of $B_{d,s}^0 \to J/\psi f_0(\to K^+ K^-)$ are also predicted in this work.
These given predictions about the $B_{d,s}^0 \to J/\psi \sigma(f_0)$ decays
await the future examinations with high precision.

\section*{acknowledgments}

X.L. thanks Professor Hai-Yang~Cheng and Professor Hsiang-nan~Li for helpful
discussions. The authors are very grateful to Professor Sheldon Stone for his
enlightening discussions and valuable comments on the manuscript.
This work is supported in part by the National Natural
Science Foundation of China under Grants  No.~11575151,  No.~11775117,
 No.~11875033,  No.~11705159,  No.~11447032 and  No.~11765012,
by the Qing Lan Project of Jiangsu Province (Grant  No.~9212218405),
by the Natural Science Foundation of Shandong Province (Grants  No.~ZR2016JL001,
 No.~ZR2018JL001,   No.~ZR2019JQ04), and by the Research Fund of Jiangsu Normal
University (Grant  No.~HB2016004).

%%%=================================================
%%%%%%%%%%%%%%%%       Apendix      ****************
%%%=================================================

\begin{appendix}
\section{\boldmath Effective Wilson Coefficients }\label{sec:app1}

As was pointed out in Ref.~\cite{Liu:2010zh}, for these considered
$B_{d,s}^0 \to J/\psi \sigma(f_0)$ decays, only the vertex corrections
%will
contribute at the currently known NLO level, in which
their effects can be absorbed into the Wilson
coefficients associated with the factorizable emission
contributions~\cite{Chay:2000xn,Cheng:2000kt},
\beq
\tilde{a}_2&=&C_1+\frac{C_2}{N_c}+\frac{\alpha_s}{4\pi}\frac{C_F}{N_c}C_2
\left(-18+12\ln\frac{m_b}{\mu}+f_I^0\right)\;,
\label{eq:a2}\\
\tilde{a}_3&=&C_3+\frac{C_4}{N_c}+\frac{\alpha_s}{4\pi}\frac{C_F}{N_c}C_4
\left(-18+12\ln\frac{m_b}{\mu}+f_I^0\right)\;,\label{eq:a3}\\
\tilde{a}_5&=&C_5+\frac{C_6}{N_c}+\frac{\alpha_s}{4\pi}\frac{C_F}{N_c}C_6
\left(6-12\ln\frac{m_b}{\mu}-f_I^0\right)\;,
\label{eq:a5}\\
\tilde{a}_7&=&C_7+\frac{C_8}{N_c}+\frac{\alpha_s}{4\pi}\frac{C_F}{N_c}C_8
\left(6-12\ln\frac{m_b}{\mu}-f_I^0\right)\;,\label{eq:a7}\\
\tilde{a}_9&=&C_9+\frac{C_{10}}{N_c}+\frac{\alpha_s}{4\pi}\frac{C_F}{N_c}C_{10}
\left(-18+12\ln\frac{m_b}{\mu}+f_I^0\right)\;,\label{eq:a9}
\eeq
with the function $f_I^0$,
\beq
f_I^0 &=& f_I + g_I (1-z) \;,
 \label{eq:fIh}
\eeq
where $z\equiv r^2_{d,s}=m^2_{J/\psi}/m^2_{B_{d,s}^0}$ and the functions $f_I$ and $g_I$ read as~\cite{Cheng:2000kt}
\beq
f_I &=& \frac{2\sqrt{2N_c}}{f_{J/\psi}}
\int^1_0 dx_2\;\phi_{J/\psi}^L(x_2)\Biggl\{ \frac{2z x_2 }{
1-z(1-x_2)}+\left(3-2x_2-8x_2^2\right)\frac{\ln x_2}{1-x_2}  \non
 && +\left(-\frac{3}{1-z x_2}+\frac{1+8 x_2}{1-z(1-x_2)}
 -\frac{2z x_2}{[1-z(1-x_2)]^2}\right)z x_2\ln z x_2  \non
 && +\left(3(1-z)+2z x_2-8z x_2^2+\frac{2z^2x_2^2}{1-z(1-x_2)}\right)
 \frac{\ln (1-z)-i\pi}{1-z(1-x_2)}\Biggr\} \;,
\label{eq:fI}
\eeq
and
 \beq
g_I &=& \frac{2\sqrt{2N_c}}{f_{J/\psi}}
\int^1_0 dx_2\;\phi_{J/\psi}^L(x_2)\Biggl\{\frac{4x_2(2x_2-1)}{
(1-z)(1-x_2)}\ln x_2+ \frac{zx_2} {[1-z(1-x_2)]^2}\ln (1-z)  \non
 && + \Biggl(\frac{1} {(1-zx_2)^2}- \frac{1} {[1-z(1-x_2)]^2}
 -\frac{8x_2} {(1-z)(1-zx_2)}  \non
&& + \frac{2(1+z-2zx_2)} {(1-z)(1-zx_2)^2}\Biggr)zx_2\ln
zx_2-i\pi\,{zx_2\over [1-z(1-x_2)]^2}\Biggr\},
 \label{eq:gI}
 \eeq
respectively.

\end{appendix}

%%%%%%%%%%%%%%%%%%%%%%%%%%%%%%%%%%%%%%%%%%%%%%%%%%%%%%%%%%
%%%%%%%   References  %%%%%%%%
%%%%%%%%%%%%%%%%%%%%%%%%%%%%%%%%%%%%%%%%%%%%%%%%%%%%%%%%%%

%\end{CJK*}
\end{document}